\documentclass[traditabstract]{aa} 

\usepackage{graphicx, times, txfonts, float, rotating, color, url, lscape, bm}

\newcommand{\less}{\raisebox{-1.1mm}{$\stackrel{<}{\sim}$}} 
\newcommand{\more}{\raisebox{-1.1mm}{$\stackrel{>}{\sim}$}} 
\newcommand{\msol}{\mbox{M$_{\odot}$}} 
\newcommand{\rsol}{\mbox{R$_{\odot}$}} 
\newcommand{\rstar}{\mbox{R$_{\star}$}} 
\newcommand{\msolyr}{{M$_{\odot}$}\,yr$^{-1}$} 

\newcommand{\lsol}{\mbox{L$_{\odot}$}}

\newcommand{\ks}{km s$^{-1}$} 
 
\newcommand{\mum}{$\mu$m} 

\newcommand{\G}{{\it Gaia}}

\begin{document} 
 
\title{Analysing the spectral energy distributions of Galactic classical Cepheids
\thanks{
Tables~\ref{Tab-Targets} and \ref{Tab:App} are available in electronic form at the CDS via 
anonymous ftp to cdsarc.u-strasbg.fr (130.79.128.5) or via 
http://cdsweb.u-strasbg.fr/cgi-bin/qcat?J/A+A/. 
Table~\ref{App:Fit} and Figure~\ref{Fig:App} are available in the on-line edition of A\&A. 
}
}  
 
\author{ 
M.~A.~T.~Groenewegen
}

\institute{ 
Koninklijke Sterrenwacht van Belgi\"e, Ringlaan 3, B--1180 Brussels, Belgium \\ \email{martin.groenewegen@oma.be}
} 
 
\date{received: ** 2019, accepted: 11-01-2020} 
 
\offprints{Martin Groenewegen} 
 
 
\abstract
{
%
Spectral energy distributions (SEDs) were constructed for a sample of 477 classical cepheids (CCs); 
including stars that have been classified in the literature as such but  are probably not.
The SEDs were fitted with a dust radiative transfer code. Four stars showed a large mid- or far-infrared excess and the
fitting then included a dust component. These comprise the well-known case of 
RS Pup, and three stars that are (likely) Type-II cepheids (T2Cs), AU Peg, QQ Per, and FQ Lac. The infrared (IR) excess in FQ Lac is
reported for the first time in this work.

The remainder of the sample was fitted with a stellar photosphere to derive the best-fitting luminosity and effective temperature.
Distance and reddening were taken from the literature. The stars were plotted in a Hertzsprung-Russell diagram (HRD) and compared
to evolutionary tracks for cepheids and theoretical instability strips. For the large majority of stars, the position
in the HRD is consistent with the instability strip for a CC or T2C. About 5\% of the stars are outliers in the sense that they are much hotter
or cooler than expected.
A comparison to effective temperatures derived from spectroscopy suggests in some cases that the photometrically derived
temperature is not correct and that this is likely linked to an incorrectly adopted reddening.

Two three-dimensional reddening models have been used to derive alternative estimates of the reddening for the sample.
There are significant systematic differences between the two estimates with a non-negligible scatter.

In this work the presence of a small near-infrared (NIR) excess, as has been proposed in the literature for a few well-known cepheids,
is investigated.
Firstly, this was done by using a sample of about a dozen stars for which a mid-infrared spectrum is available. This data is particularly constraining 
as the shape of the observed spectrum should match that of the photosphere and any dust spectrum, both dust continuum and any
spectral features of, for example, silicates or aluminium oxide. This comparison provides constraints on the dust composition, in agreement
with a previous work in the literature.
Secondly, the SEDs of all stars were fitted with a dust model to see if a statistically significant better fit could be obtained.
The results were compared to recent work.
Eight new candidates for exhibiting a NIR excess are proposed, solely based on the photometric SEDs.
Obtaining mid-infrared spectra would be needed to confirm this excess.
Finally, period-bolometric luminosity and period-radius relations are presented for samples of over 370 fundamental-mode CCs.
}


\maketitle

\section{Introduction}
\label{S-Int}

Classical Cepheids (CCs) are considered an important standard candle because they are
bright and, thus, they comprise a link between the distance scale in the nearby
universe and that further out via those galaxies that contain both Cepheids and SNIa 
(see  \citealt{Riess19} for a determination of the Hubble constant to 1.9\% precision, taking
into account the new 1.1\% precise distance to the Large Magellanic Cloud from \citealt{Pietrzynski19}).

It is therefore not surprising that the {\it Gaia} 2nd data release (GDR2, \citealt{GDR2Sum}) spurred a number of studies on the
CCs listed in the GDR2 and on the period-luminosity ($PL$) relation.
\citet{RiessGDR2} analysed a sample of 50 CCs. They derived a parallax zeropoint offset of $-0.046 \pm 0.013$ mas, compared to the
$-0.029$ mas derived for quasars by \citet{Lindegren18} and concluded that the need to independently determine the  parallax zeropoint
offset largely counters the higher accuracy of the parallaxes in determining an improved zeropoint of the $PL$-relation.
\citet{Ripepi_GDR2} re-classified all 2116 stars reported by \citet{Clementini19} to be Cepheids in the Milky Way (MW).
In total 1257 stars were classified as Cepheids (including 575 CCs pulsating in the fundamental mode (FU),
108 anomalous Cepheids (ACEP), and 336 Type-II Cepheids (T2C)). Period-Wesenheit relations in the {\it Gaia} bands were presented.
Assuming a canonical distance modulus to the LMC of 18.50, a {\it Gaia} parallax zeropoint offset of $\sim -0.07$ to $-0.1$~mas was found.
\citet{Gr_GDR2} (hereafter G18) started from an initial sample of 452 Galactic CCs with accurate [Fe/H] abundances from
spectroscopic analysis. Based on parallax data from \G\ DR2, supplemented with accurate non-\G\ parallax data when available,
a final sample of about 200 FU mode Cepheids with good astrometric solutions was retained to derive PL and
period-luminosity-metallicity ($PLZ$) relations. 
The influence of a parallax zeropoint offset on the derived $PL(Z)$ relation is large and make that the 
current GDR2 results do not allow to improve on the existing calibration of the relation, or on the distance to the LMC
(as also concluded by \citealt{RiessGDR2}). The zeropoint, the slope of the period dependence, and the metallicity dependence
of the $PL(Z)$ relations are correlated with any assumed parallax zeropoint offset.

Based on a comparison for nine CCs with the best non-\G\ parallaxes (mostly from {\it HST} data) a 
parallax zeropoint offset of $-0.049 \pm 0.018$~mas is derived, which is consistent with other values that appeared in the literature
after the release of GDR2,
from RGB stars using {\it Kepler} and {\it APOGEE} data  (about $-0.053$ mas, \citealt{Zinn18}), 
a sample of $\sim 150$ eclipsing binaries ($-0.082 \pm 0.033$ mas, \citealt{Stassun18}),
a sample of 50 CCs ($-0.046 \pm 0.013$ mas, \citealt{RiessGDR2}), 
140-300 RR Lyrae stars ($\sim -0.056$ mas, \citealt{Muraveva18}; $-0.042 \pm 0.013$ mas, \citealt{Layden19}),
a sample of 34 stars with VLBI astrometry ($-0.075 \pm 0.029$ mas, \citealt{Xu19}),
a sample of about seven million objects with a radial velocity (RV) in \G\ ($\sim -0.054$ mas,  \citealt{Schonrich19})
a sample of $\sim 250~000$ stars from APOGEE ($-0.0523 \pm 0.020$ mas, \citealt{LeungBovy19}),
a sample of $\sim 27~000$ Red Clump stars selected from APOGEE ($-0.048 \pm 0.01$ mas, \citealt{ChanBovy19}).
These values are mostly all-sky averages, but when sufficient data is available it is clear that the parallax zeropoint offset depends on
position on the sky, magnitude, and colour \citep{Zinn18, Khan19, LeungBovy19, ChanBovy19}.

The analysis by \citet{Ripepi_GDR2} on the classification of CCs addresses one of the issues that also affected the analysis in G18.
The classification as CCs was taken from the literature in that paper and the origin of this
classification is sometimes hard to trace. In addition, some stars have alternative classifications reported in the literature.
It is clear that the most accurate determination of $PL$- or period-radius ($PR$) relations would benefit from a `clean' sample.

To address this issue the spectral energy distributions (SEDs) of the sample in G18 are constructed in the present paper,
and fitted with model atmospheres (and a dust component if needed). For a given distance and reddening this results
in the absolute luminosity and (photometric) effective temperature. Placing the objects in the Hertzsprung-Russell diagram (HRD)
and comparing the location to theoretical instability strips (ISs) and evolutionary tracks may show whether the derived stellar
parameters are consistent with the variability classification as CCs.
In addition, such a procedure may reveal stars whose SEDs are not well fitted by a stellar atmosphere, and that show the
presence of infrared emission, such as observed and postulated in a number of well-known CCs
($\delta$ Cep, $\eta$ Aql, X Sgr, T Mon, $l$ Car, Y Oph,
see \citealt{Merand2005,Merand2006,Kervella2006,Merand2007,Gallenne13b,Merand2015,Breitfelder16}), or known to occur in some T2C
(in particular RV Tau [RVT] variables, but recently also seen in lower-luminosity W Vir stars, see \citealt{Kamath2016,GrJu17a} and
references therein) that could be misclassified as CCs.

The paper is structured as follows. In Section~\ref{S-Sam} the sample of stars is briefly described.
Section~\ref{S-SED} describes the construction of the SEDs and the model fitting.
Section~{\ref{S-RES}} presents the results of the calculations in various subsection.
A brief discussion and summary concludes the paper.

\section{The sample} 
\label{S-Sam}

The sample studied here is the sample of 452 stars considered in G18 along with 25 additional stars, as described below.
G18 compiled a list all CCs with individually determined accurate iron abundances from high-resolution spectroscopy.
Some of the stars in the sample had alternative classifications in the literature or were even unlikely to be CCs but
they are retained here for completeness. 
Since then, \citet{Luck18} published a list of abundances and parameters for 435 Cepheids, 20 of which were not in the G18 sample.
In addition, \citet{Inno19} (hereafter I19) recently determined the metallicity of five CCs in the inner disk of our Galaxy that are of interest.
The sample considered in this paper is therefore 477 objects.
The basic information for this sample is listed in Table~\ref{Tab-Targets}. The pulsation type listed in Col.~2, the period  (Col.~4)
and  the $E(B-V$ values  (Col.~5)
are taken from G18 for the first 452 stars (based on compilations in the literature), \citet{Luck18} for the next 20 stars and
I19 for the stars in the direction of the inner disk (with identifier ID 1-5 following the nomenclature in Inno et al.).
In the case of the inner disk cepheids the $A_{K_{\rm s}}$ values from col.~3 in Table~3 in I19 based on the \cite{Cardelli89} reddening law were taken, converted
  to $A_{\rm V}$ using $A_{K_{\rm s}}/A_{\rm V} = 0.114$ and then converted to $E(B-V)$ using a specific reddening of 3.1.
The pulsation type listed in Col.~3 is from the independent classification by \citet{Ripepi_GDR2}.
The adopted distance ($d$) is listed in Col.~6 based on the reference in Col.~7. When available this is based on parallax data,
otherwise it is the distance quoted in the relevant papers, typically based on a $PL$-relation. In the case of a \G\ parallax,
a parallax zero-point offset of $-0.043$ mas was adopted, following G18. The exact value of this offset, or the adopted
distance in general, is not so crucial as it was in G18 or in other papers that aim to improve the $PL$-relation.
The derived luminosities will scale with $d^2$, and the derived effective temperatures are independent of the adopted distance.
To give some feeling of the distance and the possible range in distances, Col.~9 reports the distance and error
from \citet{BJ18} based on a Bayesian analysis taking into account a three-dimensional (3D) model of the Galaxy as prior
and using a parallax zero-point offset of $-0.029$ mas.

In general, the distances are in agreement within the margin of error.
In only three cases do the adopted distance and the distance from \citet{BJ18} differ by more than 3$\sigma$ and would this difference
in distance lead to a difference in luminosity larger than a factor of three.
They are
EF Tau, RW Cam, TX Del, and V1359 Aql. Only EF Tau is classified as a CC, while the others are not.

\longtab{
  \begin{landscape}
    \small
\setlength{\tabcolsep}{1.3mm}
\begin{longtable}{rlrrcrrrccclll}
  \caption{\label{Tab-Targets} Sample of stars.} \\
  \hline  \hline
Name      & Type     &  Type  &  Period    &   $E(B-V)$        &  $d$   & Ref & $\sigma_{\rm d}$ & distance & Luminosity  &  $T_{\rm eff}$   & $\theta$ & Remarks  \\ 
          &          &        &   (d)      &    (mag)          &  (kpc)  &     &  (kpc) &   (kpc)  &   (\lsol)    &    (K)        &    (mas)  &        \\ 
\hline
\endfirsthead
\caption{continued.}\\
\hline\hline
Name      & Type     &  Type  &  Period    &   $E(B-V)$        &  $d$    & Ref & $\sigma_{\rm d}$ & distance & Luminosity  &  $T_{\rm eff}$   & $\theta$ & Remarks  \\ 
          &          &        &   (d)      &    (mag)          &  (kpc)  &     &  (kpc)        &   (kpc)  &   (\lsol)    &    (K)        &    (mas)  &        \\ 
\hline
\endhead
\hline
\endfoot
   AA Gem          & DCEP     & DCEP\_F     &  11.302 & 0.34 $\pm$ 0.04 &  3.400 &  9 &  0.829 &  4.230$^{+1.123}_{-0.770}$ &   3400.0 $\pm$   122.7 & 5500 $\pm$  302 & 0.176 $\pm$ 0.018 &  \\      
   AA Mon          & DCEP     &            &   3.938 & 0.77 $\pm$ 0.02 &  3.922 &  1 &  0.709 &  4.014$^{+0.781}_{-0.575}$ &    922.8 $\pm$    33.6 & 5500 $\pm$  177 & 0.079 $\pm$ 0.005 &  \\      
   AB Cam          & DCEP     &            &   5.788 & 0.62 $\pm$ 0.04 &  4.200 &  9 &  0.966 &  5.150$^{+1.285}_{-0.897}$ &   1463.5 $\pm$    79.3 & 5875 $\pm$  277 & 0.082 $\pm$ 0.008 &  \\      
   AC Mon          & DCEP     & DCEP\_F     &   8.014 & 0.51 $\pm$ 0.03 &  2.400 &  9 &  0.400 &  3.327$^{+0.596}_{-0.445}$ &   1991.6 $\pm$    42.2 & 5375 $\pm$  168 & 0.200 $\pm$ 0.012 &  \\      
   AD Cam          & DCEP     & DCEP\_F     &  11.261 & 0.87 $\pm$ 0.01 &  4.600 &  9 &  0.756 &  4.463$^{+0.785}_{-0.592}$ &   2048.8 $\pm$    87.0 & 5125 $\pm$  253 & 0.116 $\pm$ 0.011 &  \\      
   AD Cru          & DCEP     & DCEP\_F     &   6.398 & 0.64 $\pm$ 0.01 &  2.994 &  1 &  0.394 &  3.118$^{+0.435}_{-0.343}$ &   1881.9 $\pm$    93.2 & 5500 $\pm$  237 & 0.149 $\pm$ 0.013 &  \\      
   AD Gem          & DCEP     & DCEP\_F     &   3.788 & 0.21 $\pm$ 0.05 &  2.500 &  9 &  0.673 &  5.192$^{+1.484}_{-1.048}$ &    966.0 $\pm$    32.0 & 6000 $\pm$  306 & 0.107 $\pm$ 0.010 &  \\      
   AD Pup          & DCEP     & DCEP\_F     &  13.596 & 0.36 $\pm$ 0.02 &  4.100 &  9 &  0.946 &  5.769$^{+1.436}_{-1.013}$ &   4650.8 $\pm$   356.8 & 5500 $\pm$  302 & 0.171 $\pm$ 0.018 &  \\
\end{longtable}
\tablefoot{
  Column~1. The variable star name or identifier. The first 452 objects are from \citet{Gr_GDR2}, the last 25 (BE Pup and later) represent the stars added to the sample (see main text).
  ASAS1810 is short for  181024-20, and ASAS1713 is short for ASAS 171305-43.
%
Column~2. The classification of the variability, see \citet{Gr_GDR2} for the first 452 stars, and the main text for the added stars.
Column~3. The classification by  \citet{Ripepi_GDR2} who re-classified the Cepheids in the GDR2. Multimode cepheids are labelled with "\_MU".
Column~4. The pulsation period in days. 
Column~5. Reddening value $E(B-V)$ with error bar. From \citet{Gr_GDR2}, I19 (see main text for details), and \citet{Luck18} (except V1206 Cas, V701 Car and V898 Cen from \citealt{Stevens17}).
Column~6. Adopted distance, with reference (Col.~7).
(1) parallax from \citet{GDR2Sum} with additional criteria and a parallax zeropoint offset applied (see main text), 
(2) parallax from \citet{vanL08},  
(3) parallax from \citet{vanL07},  
(4) parallax from \citet{Benedict07}, 
(5) parallax from \citet{Riess14},
(6) parallax from \citet{Riess18}. For references 1-6 the distance is taken as 1/parallax.
(7) \citet{Gallenne2018} 
(8) \citet{Inno19},      
(9) \citet{Melnik15},
(10) \citet{Acharova2012}, 
(11) \citet{Genovali2014},  
(12) \citet{Kashuba16}, 
(13) \citet{Andrievsky2016},  
(14) \citet{MarAnd15},      
(15) \citet{Luck18},        
(16) \citet{Luck14}.        
Column~8.  Adopted error in the Distance.
Column~9. Distance with error bar as given by \citet{BJ18}.
Column~10. Luminosity with error bar, for the adopted distance.
Column~11. Effective temperature with error bar.
Column~12. Angular diameter with error bar.
Column~13. Any remarks.
Angular diameters from the literature are referenced as follows:
(a)= \citet{Merand2006},  (b)=\citet{Jacob08}, (c)= \citet{Kervella2004}, (d)= \citet{Merand2005},  (e)= \citet{Gallenne2012}, (f)= \citet{Davis09}, (g)= \citet{Kervella04},
(h)= \citet{Kervella2006}, (i)= \cite{Kervella2017}, (j)= \citet{Gallenne13b}, (k)= \citet{Gallenne13}, (l)= \citet{Merand2007},
(m)= \citet{Gallenne19}, indicated is the mean value over the listed number of epochs.
}
\end{landscape}
}

\section{Photometric data and SED fitting} 
\label{S-SED}

The spectral energy distributions (SEDs) are constructed using photometry retrieved mostly, but not exclusively, 
via the VizieR web-interface\footnote{\url{http://vizier.u-strasbg.fr/viz-bin/VizieR}}.
Given the variability of the sources the aim is to use, as much as possible, magnitudes (and their error bars) at mean light.
The optical data comes from GDR2 ($G$, $B_{\rm p}$ and $R_{\rm p}$), 
\citet{Berdnikov2008}, \citet{Berdnikov2015}, \citet{Mermilliod1997}, \citet{Droege2006},
APASS (AAVSO Photometric All Sky Survey DR9, \citealt{Henden16}),
and data available throught the McMaster database\footnote{\url{https://www.physics.mcmaster.ca/Cepheid/}} initiated by Dr. Welch.
Attention is given to include Walraven photometry from \citet{Walraven64} and \citet{Pel76} as this provides a valuable
source of photometric data in the blue part of the spectrum.
Also GALEX data from \citet{Bianchi17} is considered.
In some case the individual epoch photometry is fitted with the code {\sc Period04} \citep{Period04} to obtain
the mean magnitudes and error bar.

The near-infrared (NIR) photometry is more heterogeneous as it comes from a variety of sources, using different photometric systems and 
ranges from intensity-mean magnitudes from well sampled light curves to single-epoch photometry in some cases.
Details are given in G18, but in brief, mean magnitudes are taken from 
\citet{MP11}  (converted to the 2MASS system), 
SAAO-based photometry (mainly \citealt{Laney1992}, and Laney (priv. comm.) as quoted in \citet{Genovali2014}, and \citealt{Feast2008},  
and CIT-based photometry from \citet{Welch1984} and \citet{Barnes1997}, converted to the 2MASS system. 
Additional single-epoch photometry is taken from \citet{McGonegal83}, \citet{Welch1984}, \citet{Schechter1992}, DENIS,
2MASS, 2MASS 6X, IRSF \citep{Kato_IRSF}.
Single-epoch NIR data is available from fourth data release of the VVV survey \citep{Minniti2010}\footnote{see \url{http://horus.roe.ac.uk/vsa/index.html}}
for ID~1,2, and 5.

At longer wavelengths generally no light-curve averaged mean magnitudes exist, but the photometric pulsation amplitudes decrease with
increasing wavelength and so the effect of the variability on the derived luminosity will be less.
An exception is \citet{Monson12} who present intensity-averaged magnitudes in the {\it Spitzer} IRAC 3.6 and 4.5~$\mu m$ bands
for 37 cepheids.
\citet{Marengo10} give single-epoch {\it Spitzer} data in all four IRAC bands and MIPS 24 and 70~$\mu m$ for 29 cepheids
(only for nine stars in [70]).
Additional single-epoch IRAC and MIPS photometry is available in the GLIMPSE \citep{GLIMPS09} and MIPSGAL \citep{Gutermuth15} catalogues.
Single-epoch Akari data is available at 9 and 18~$\mu m$ from \cite{AKIRC10}.
Akari data at longer wavelengths (FIS, \citealt{AkFIS07}) is available for two objects (BQ Ser and V1344 Aql) but is unreliable.
Averaged WISE data is available for the majority of objects \citep{Cutri_Allwise}.
Finally, data from the IRAS {\it Point Source Catalog} (PSC, \citealt{Beichmann85}), the COBE-DIRBE PSC \citep{COBEDIRBE}
and narrow-band filter data from \citet{GalVISIR} are added.

The smallest number of photometric data points over the different filters is nine (for two stars).
On the other hand there are fifteen stars with 40 or more data points. 
The median number of data points is twenty-five.

Mid-IR (MIR) spectra are available for more than a dozen stars. This is particularly useful data in the detection of IR excess.
{\it Spitzer} IRS spectra are retrieved using the CASSIS tool\footnote{\url{https://cassis.sirtf.com/}}
(Combined Atlas of Sources with Spitzer IRS spectra, \citealt{Lebouteiller11}) for
AY Cen,  $\eta$ Aql,  S TrA, SU Cyg, V Cen, V1334 Cyg, $\zeta$ Gem, 
Polaris, $\delta$ Cep, $l$ Car, and
RS Pup.
{\it IRAS} LRS spectra are retrieved for $\beta$ Dor and V382 Car (as well as Polaris, $l$ Car, and $\eta$ Aql, but for
which the higher quality IRS spectra is used) using the interface provided
by Dr. Volk\footnote{\url{http://isc83.astro.unc.edu/iraslrs/getlrs_test.html}}.
In addition, spectra are available for T Mon and X Sgr from the MIDI instrument \citep{Gallenne13b}.
All these objects are explicitly discussed in Sect.~\ref{S-EXC}.

The SEDs are fitted with More of DUSTY (MoD, \cite{Gr_MOD})\footnote{http://homepage.oma.be/marting/codes.html}
which uses a slightly updated and modified version of the DUSTY dust radiative transfer (RT) code \citep{Ivezic_D} as
  a subroutine within a minimisation code. 

Input to the model are the distance, reddening, a model atmosphere, and the absorption and scattering coefficients of any dust component.
For a given set of observed photometric data and spectra (and visibility data, and 1D intensity profiles) 
the program determines the best fitting luminosity ($L$), dust optical depth ($\tau$, at 0.55 $\mu$m), 
dust temperature at the inner radius ($T_{\rm c}$), and slope of the density profile ($\rho \sim r^{-p}$)
by minimising a $\chi^2$ based on every available photometric and spectroscopic datapoint and its error.
Any of these parameters can also be fixed.

The model fluxes are reddened to be compared to the observations using the input value for $E(B-V)$, a specific reddening of 3.1, and the
reddening law from \cite{Cardelli89} and \cite{ODonnell94} from the UV to the NIR and with the MIR silicate extinction curve from the
Local ISM model in \citet{Chiar06}.
The comparison to the observed magnitudes is done by convolving the model SED with a large number of photometric filters with the appropriate zeropoints.

The SEDs are fitted under the assumption of being representative of a single star.
Any unresolved binary will influence the photometry depending on the luminosity ratio and difference in 
spectral type and hence the resulting effective temperature and luminosity (see Sect.~\ref{S-BIN} for an estimate of the effect)
MARCS model atmospheres are used as input \citep{Gustafsson_MARCS} with solar metallicity and a $\log g= 2$.  
The model grid is available at 250~K intervals for the effective temperature 
range of interest, and adjacent model atmospheres are used to interpolate models at 125~K intervals, 
which reflects better the accuracy in $T_{\rm eff}$ that can be achieved.
Most stars have no dust and are best represented by a `naked' star. In those cases, the dust optical depth is 
fixed to a very small number ($10^{-5}$, and $T_{\rm c}$ and $p$ are also fixed to standard values of 1000~K and 2, respectively).
For every model atmosphere (that is, $T_{\rm eff}$) a best-fitting luminosity (with its [internal] error, based on the covariance matrix)
is derived with the corresponding reduced $\chi^2$ ($\chi_{\rm r}^2$) of the fit. 
The model with the lowest $\chi_{\rm r}^2$ then gives the best-fitting effective temperature. 
Considering models within a certain range above this minimum $\chi_{\rm r}^2$ then gives the error in the 
effective temperature and luminosity. For the luminosity this error is added in quadrature to the internal error in luminosity.

For some stars a better fit is achieved by adding a dust component. 
The BIC (Bayesian information criterion, see \cite{Schwarz1978}) is used to verify if the lower $\chi^2$ that 
is obviously obtained when adding additional parameters is, in fact, statistically significant.

In the next section the results of the various calculations are presented.

\section{Results}
\label{S-RES}

\subsection{Mid-IR and Far-IR excess}
\label{S-MIR}

A visual inspection of the SEDs revealed four stars that evidently showed an IR excess.
We note that this large excess is different from the excess of order a few percent that is claimed in a number of CCs (see introduction) and whose nature
is explicitly investigated in Sect.~\ref{S-EXC}.

One of the four is RS Pup and its excess in the far-IR is long known \citep{Gehrz70,McAlary86,Deasy1986}.
The IRS spectrum that is used in the SED fitting is that of the emission close to the star and does not include the extended emission.
In the SED fitting the part of the spectrum beyond 20~$\mu m$ is excluded not no influence the fitting of the extended dust component.

\begin{figure}

\begin{minipage}{0.45\textwidth}
\resizebox{\hsize}{!}{\includegraphics{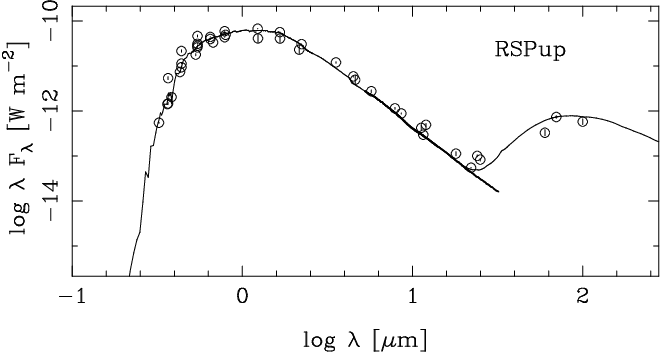}}
\end{minipage}

\begin{minipage}{0.45\textwidth}
\resizebox{\hsize}{!}{\includegraphics{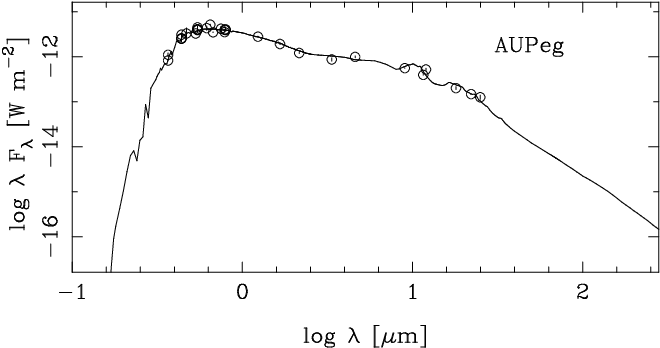}}
\end{minipage}

\begin{minipage}{0.45\textwidth}
\resizebox{\hsize}{!}{\includegraphics{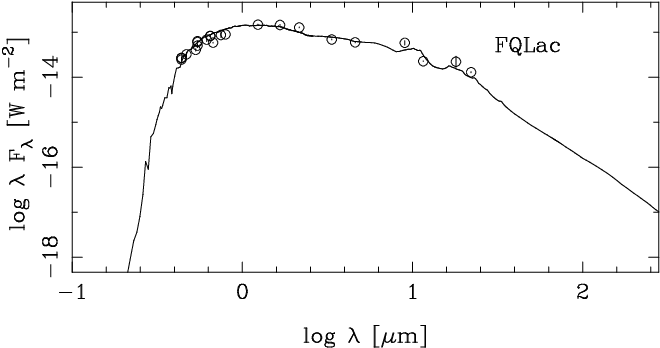}}
\end{minipage}

\begin{minipage}{0.45\textwidth}
\resizebox{\hsize}{!}{\includegraphics{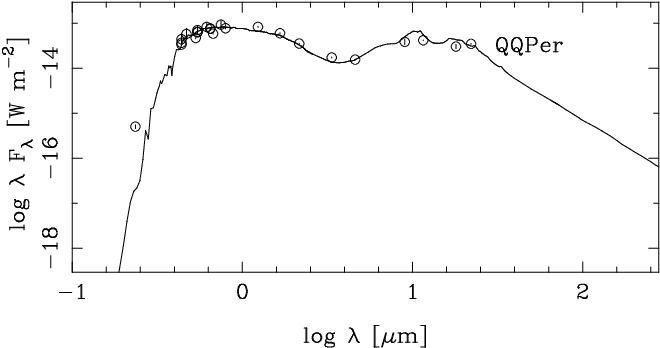}}
\end{minipage}

\caption{Best fit models of RS Pup, and the three (likely) T2Cs that show near- and mid-IR excess.
Errorbars are plotted and are typically the size of the plot symbol.
}
\label{Fig:T2C}
\end{figure}

The other three stars are AU Peg, FQ Lac, and QQ Per. The SEDs of these stars show an near- and mid-IR excess that is typical of that of RVT stars and also
recently seen in a number of lower-luminosity W Vir stars in the Magellanic Clouds \citep{Kamath2016,GrJu17a}.
The adopted classification in G18 is CWB, CEP:?, CEP?
, respectively, the classification in \citet{Ripepi_GDR2} is
BLHER, Fundamental mode CC, and WVIR, respectively (see Table~\ref{Tab-Targets}).
The fitting of the SED (also in the case of RS Pup) is performed as outlined in \citet{GrJu17a} and includes a dust component (see \citealt{GrJu17a} for details).
The best-fit SEDs are shown in Fig.~\ref{Fig:T2C}.
It should be pointed out that the shape of the excess
points to a a disc structure rather than an expanding 
outflow, so the use of a 1-D code is limited. For the purpose of the present paper we included a realistic dust component in order to get a more realistic
estimate of the luminosity.

The dust temperature at the inner radius is found to be 46~K in the case of RS Pup, and 450-1050~K in the case of the T2Cs.
For RS Pup this is close to the value of `around 40~K' derived in \citet{Deasy1986} by  a blackbody fit to the IRAS data.
The IR excess in AU Peg was detected first by \citet{McAlary86} based on IRAS data. 
The SED of QQ Per was shown in \citet{Schmidt15MIR} and identified as having a strong IR excess, but classified as a CC. 
The IR excess in FQ Lac seems to be reported for the first time in the present work it appears.

\subsection{The standard case}
\label{S-STD}

\begin{figure}
\centering

\begin{minipage}{0.49\textwidth}
\resizebox{\hsize}{!}{\includegraphics{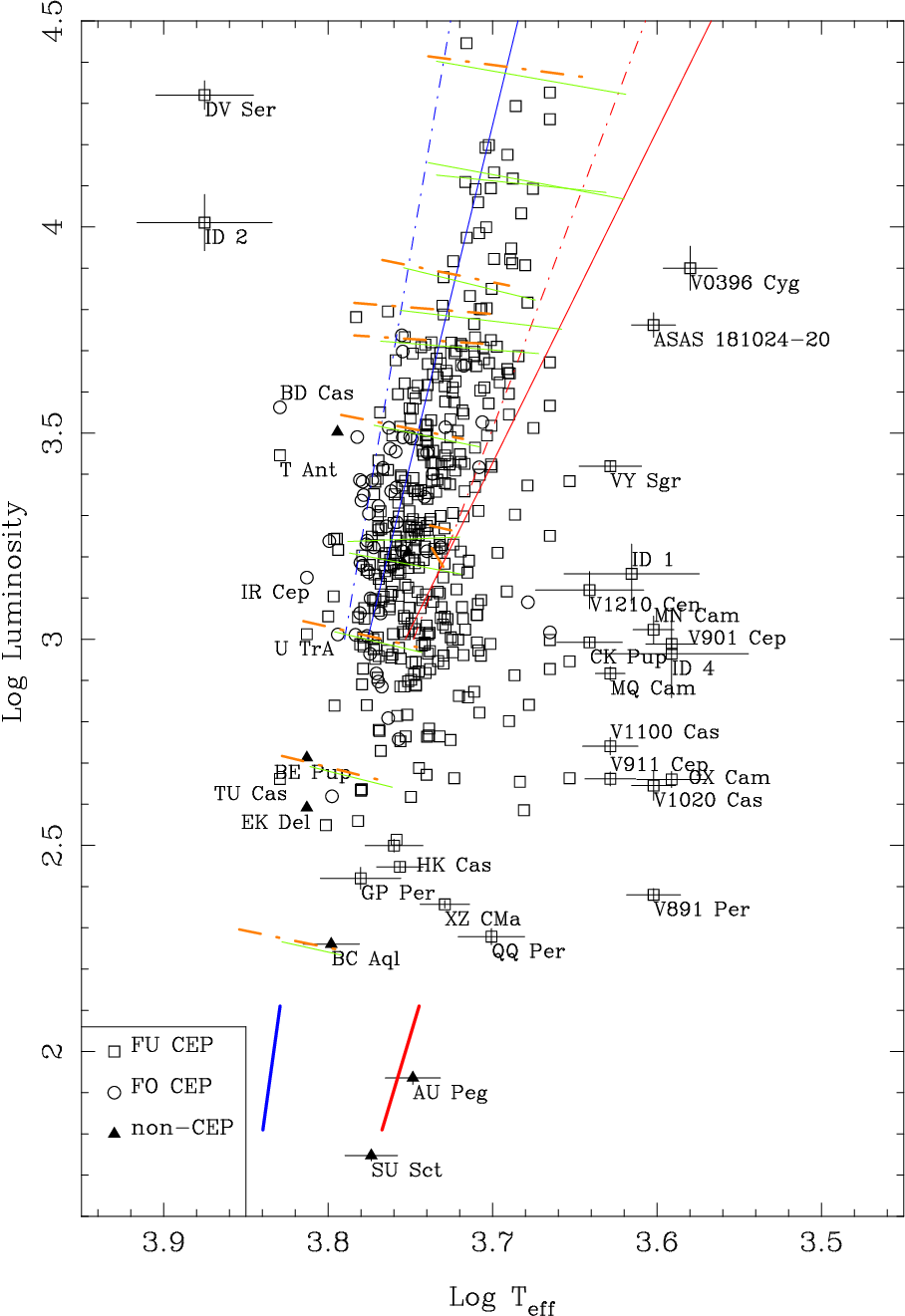}}
\end{minipage}

\caption{The Hertzsprung-Russell diagram.
  Stars located outside the bulk of objects have been labelled and are plotted with error bars.
  Blue and red lines indicate the blue and red edge of the IS.
  At $\log L \sim 2$ this is the IS for BLHER T2C; at brighter luminosities those for CCs. 
  The near horizontal lines are stellar evolution tracks of CCs of different masses.
  See the main text for details.
}
\label{Fig:HRD}
\end{figure}

In this subsection, the derived luminosities and effective temperatures are discussed in the standard case, that is fitting model atmospheres to the SEDs
without circumstellar dust, except for the four stars just discussed. The results are listed in Col.~9 and 10 of Table~\ref{Tab-Targets}.
The HRD is shown in Fig.~\ref{Fig:HRD}.
To assist in the interpretation some ISs have been plotted.
At $\log L \sim 2$ these are the blue and red edge of BL Her (T2C with period \less\ eight days) for a mass of 0.65~\msol\ \citep{DiCriscienzo_2007}.
Unfortunately, no ISs seem to be available in the literature for the brighter T2Cs, like the WVIR.
The dashed (indicating $Z= 0.008$) and full lines ($Z= 0.02$) represent the  blue and red edge for CCs from \citet{Bono_2000}.
The near horizontal lines indicate the evolutionary tracks for $Z= 0.014$ and average initial rotation rate $\omega_{\rm ini} = 0.5$ from \cite{Anderson16}.
The FO (red dot-dashed lines) and FU (green full lines) tracks are shifted by 0.01 dex in luminosity for clarity.
Increasing in luminosity they are tracks for initial mass (number of the crossing through the IS): 
3 (1), 4 (1), 5 (1), 5 (2), 5 (3), 7 (1), 7 (2), 7 (3), 9 (1), 9 (2), 9 (3), 12 (1).

The bulk of stars located between $\log L \sim 2.9-3.8$~\lsol\ would correspond to stars of initial mass $\sim 5 - 7$~\msol\ most likely in their
2nd or 3rd crossing of the IS. The evolutionary time spent in the 1st crossing is an order of magnitude shorter and this explains qualitatively
the lack of stars in the luminosity range covered by the 3 and 4~\msol\ tracks. The brightest stars in the sample would correspond to $\sim 12$~\msol\
stars during their first crossing of the IS.

The location of the majority of stars in the HRS is consistent with the location of the IS of T2C and CCs.
Error bars are plotted for some of the stars outside the bulk of objects, but they are typical for the entire sample.
Based on this there are a few stars (notably DY Ser and ID~2) that are much hotter and about two dozen stars ($\sim$ 5\% of the sample)
that are cooler than expected for a star located in the IS.
In particular for three of the five stars in the sample in the direction of the inner disk the location in the HRD appears to be inconsistent
with the IS. One obvious reason for this discrepancy is the degeneracy between interstellar reddening and the derived effective temperature.
This is explicitly investigated in the next section, and more generally in Sect.~\ref{S-EBV}.

\subsection{The role of reddening for the cepheids in the inner disk}
\label{S-IND}

\begin{figure}
\begin{minipage}{0.49\textwidth}
\resizebox{\hsize}{!}{\includegraphics{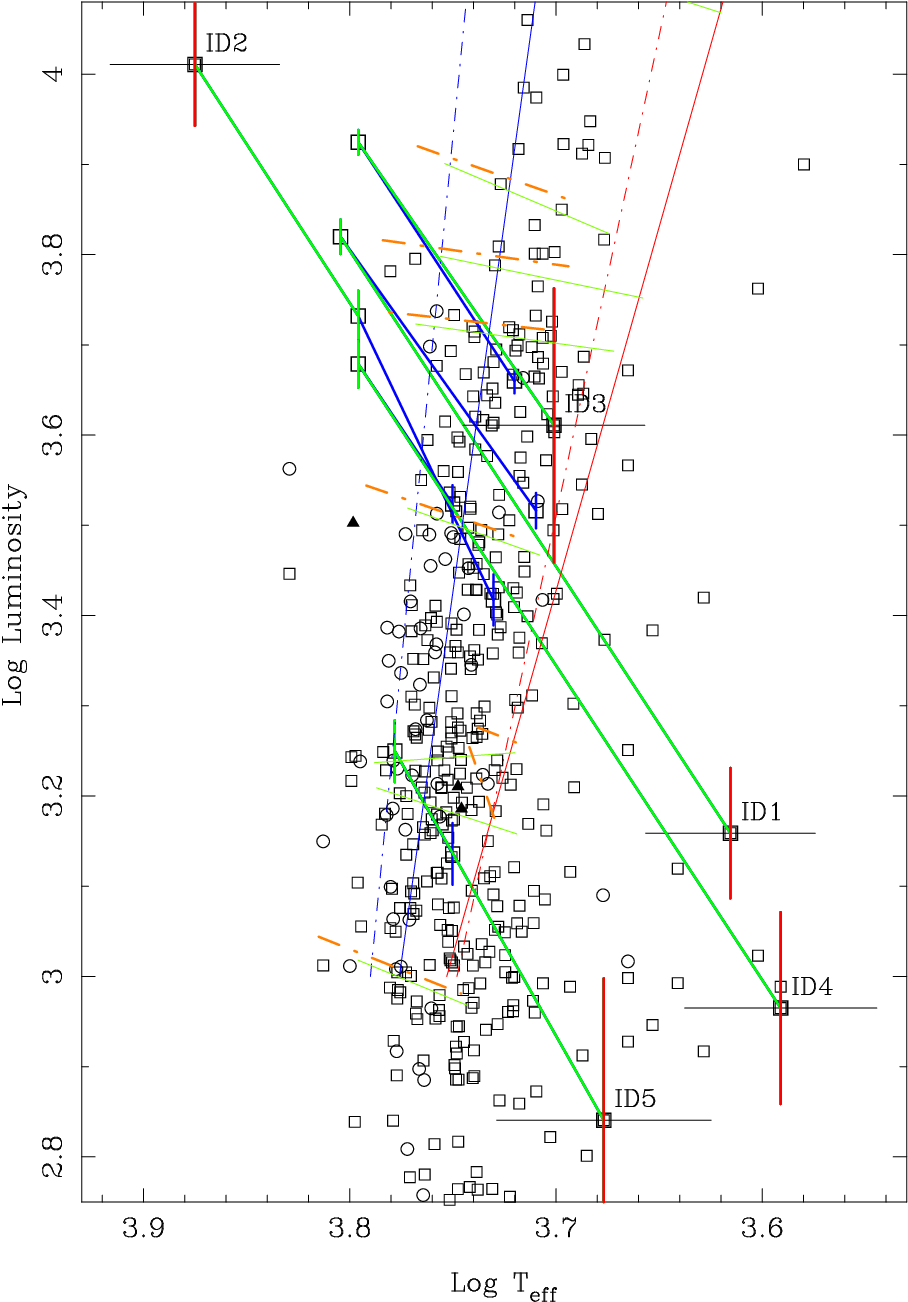}}
\end{minipage}

\caption{Hertzsprung-Russell diagram, zoomed in on the region covered by the five object in the direction of the Galactic disk.
  The labelled points with the red vertical error bars indicate the standard case (Fig.~\ref{Fig:HRD}).
  They are connected by  green lines to the points with the green vertical error bars indicating the cases where
  the effective temperature is fixed to the value determined from spectroscopy, which in turn
  are connected by blue lines to the points with blue vertical error bars indicating the cases where the effective temperature
  is fixed to a value roughly in the middle of the IS.
}
\label{Fig:HRDB}
\end{figure}

In the standard case, the interstellar reddening is fixed from I19 and the SEDs are fitted to give the best fit effective temperature and luminosity (for
the distance quoted in I19). The reddening is large and has a large formal error bar that is not explicitly considered in the fitting.
An alternative is to fix the effective temperature to the value derived spectroscopically and then determine the best-fitting luminosity and value
of $A_{\rm V}$. A third way is to fix $T_{\rm eff}$ to a value that would put the star roughly in the middle of the IS and then fit for
$L$ and $A_{\rm V}$. The results of the calculations are reported in Table~\ref{Tab:IDebv} and the HRD is shown in Fig.~\ref{Fig:HRDB}.
For ID~2, 3, 5 the temperature determination is based on a single spectrum, for ID~1 and ID~4 the average of two determinations is used.
As the temperature changes over the pulsation cycle this introduces additional uncertainty as in the case of ID~1 the two temperature determinations
differed by 700~K. The impact of a change of effective temperature is large as shown in Fig.~\ref{Fig:HRDB}, and are strongly correlated with $A_{\rm V}$
and $L$. When both the spectroscopic temperature or a temperature in the IS is chosen the resulting $A_{\rm V}$ is larger than the value adopted
in I19 for all five stars. This is an additional complication in deriving the parameters of these stars as the distances derived in I19
are based on infrared $PL$-relations that were dereddened using certain values of $A_{\rm K}$ (and also depend on the reddening law).

\begin{table*} 
\caption{CCs in the inner disk. Variation in the parameters.}

\begin{tabular}{ccrrcrrcrr} \hline \hline 

  Name &  \multicolumn{3}{c}{Photometry} & \multicolumn{3}{c}{Spectroscopy} &  \multicolumn{3}{c}{Instability strip}  \\
       &  $T_{\rm eff}$ & $A_{\rm V}$ & $L$ &  $T_{\rm eff}$ & $A_{\rm V}$ & $L$ &  $T_{\rm eff}$ & $A_{\rm V}$ & $L$  \\
       &    (K)       &  (mag)     & (\lsol) &  (K)       &  (mag)     & (\lsol) &  (K)       &  (mag)     & (\lsol) \\
\hline 

 ID 1  & 4125 $\pm$  409 &  8.6 &   1442 $\pm$   261 & 6375 & 11.8 & 6604 $\pm$ 297 & 5125 & 10.6 & 3283 $\pm$ 152 \\ 
 ID 2  & 7500 $\pm$  743 & 15.4 &  10251 $\pm$  1736 & 6250 & 14.7 & 5393 $\pm$ 364 & 5375 & 13.4 & 2614 $\pm$ 176  \\ 
 ID 3  & 5000 $\pm$  533 & 16.1 &   4081 $\pm$  1708 & 6250 & 17.1 & 8405 $\pm$ 269 & 5250 & 16.1 & 4570 $\pm$ 146 \\ 
 ID 4  & 3900 $\pm$  442 &  7.4 &    923 $\pm$   256 & 6250 & 10.8 & 4773 $\pm$ 236 & 5625 & 10.3 & 3338 $\pm$ 165  \\ 
 ID 5  & 4750 $\pm$  604 & 11.8 &    693 $\pm$   302 & 6000 & 13.9 & 1777 $\pm$ 146 & 5625 & 13.4 & 1367 $\pm$ 112 \\ 

\hline

\end{tabular} 
\label{Tab:IDebv}
\tablefoot{
Columns~2-4 give the results for the standard fitting of the SED.
The interstellar reddening is fixed and the effective temperature and luminosity are fitted.
The results are copied from Table~\ref{Tab-Targets}.
Columns~5-7 give the results when the effective temperature is fixed to the spectroscopic value.
Quoted is the effective temperature of the model atmosphere closest to it.
\citet{Inno19} quote an error of $\pm 300$~K on the spectroscopic effective temperature determination.
The luminosity and interstellar reddening are fitted.
The error on $A_{\rm V}$ is estimated to be about 1~mag.
The error on the luminosity is the formal error scaled to give a reduced $\chi^2$ of unity.
Columns~8-10 give the results when the effective temperature is fixed to a value that `by eye' would put the star
roughly in the middle of the instability strip (cf. Figure~\ref{Fig:HRD}).
The luminosity and interstellar reddening are fitted.
The error on $A_{\rm V}$ is estimated to be about 1~mag.
The error on the luminosity is the formal error scaled to give a reduced $\chi^2$ of unity.
}
  
\end{table*}

\subsection{Angular diameters}
\label{S-THETA}

Column~11 of Table~\ref{Tab-Targets} gives the predicted angular diameter with error, determined from luminosity, effective temperature, and distance.

As an aposteriori verification they are compared to observed angular diameters that are available for almost two dozen and that are listed in Col.~12.
With the exception of the data in \citet{Gallenne19}, the values represent the mean angular diameter over the light curve.
For 20 stars the predicted and observed angular diameters agree within the error bars given confidence to the fitting method and the derived parameters.
Exceptions are T Vul and the overtone pulsator AH Vel.

\subsection{Role of binarity}
\label{S-BIN}

The standard case assumes that the SED is that of a single star, the CC.
Many CCs are known to be in binaries, see, for example, the database of \citet{Szabados03}\footnote{\url{https://konkoly.hu/CEP/intro.html}}, and
the photometry extracted from the literature could be contaminated by an companion.

Three cases have been studied in detail on what the largest likely effect would be of a (probable) companion.
\citet{Kervella19CPM} looked for resolved common proper motion pairs among CCs and RR Lyrae using GDR2 data and found
27 resolved high-probability gravitationally-bound systems with CCs out of 456 stars examined.
Their Table~A1 list the {\it Gaia} photometry of the CC and the bound candidates.
Two stars are selected where this difference is smallest in the $B_{\rm p}$ band ($\sim 3.3$ mag), U Sgr and EV Sct.
In the other 25 cases this difference is much larger, up to nine magnitudes.
In fact the bound candidates are located at 25 and 72\arcsec\ away from the CC, so in reality they do not contaminate the cepheid,
but one can make the experiment if companions of this type were in fact close physical companions.

The companion to U Sgr is thought to be of spectral type A0 \citep{Kervella19CPM} and a model atmosphere of a 10~000~K star is fitted to the
{\it Gaia} $B_{\rm p}$, $G$, and $R_{\rm p}$ photometry assuming the same distance and reddening as for U Sgr. The best fit resulted in a luminosity of
$L \sim 110$~\lsol.   
The predicted magnitudes of this star were then added to those of U Sgr, and the fit of the SED is repeated.
The best-fitting luminosity is increased by 2\%, while the best-fitting effective temperature remains unchanged indicating the effect
is less than the grid interval of 125~K.

In the case of EV Sct, the spectral type of the companion is of spectral type B9 \citep{Kervella19CPM} and a model atmosphere of a 11~000~K star is fitted to 
{\it Gaia} $B_{\rm p}$, $G$, and $R_{\rm p}$ and 2MASS $JHK$ photometry.
The best fit resulted in a luminosity of $L \sim 75$~\lsol.
The predicted magnitudes of this star are again added, and the fit of the SED of EV Sct is repeated.
The results are very similar to those of U Sgr, the best-fitting effective temperature remains unchanged and the luminosity increases by 3\%.

The third case is V1334 Cyg, a system with a close companion  of spectral type B7 located at 8.5 mas \citep{Gallenne2018} that does contaminate
the photometry of the system.
As this system has all orbital parameters, masses and distance determined with high precision from a combination of (optical) interferometry
and spectroscopy \citep{Gallenne2018} it also serves as an excellent system to test the methodology to search for companions from the difference between
{\it Hipparcos} and {\it Gaia} proper motions \citep{Kervella19PMA}.
A model atmosphere of a 15~000~K star is fitted to a 2MASS $H$-band of 8.47~mag, which is based on the estimated flux-ratio of the Cepheid and the companion
($\Delta H= 3.70 \pm 0.11$) from NIR interferometry \citep{Gallenne2018}.
The predicted magnitudes of this star are subtracted, and the fit of the SED of V1334 Cyg is repeated.
In this (more uncertain) case the best fitting temperature shifts to the next point in the available grid (from 5875 to 5750~K) and the luminosity
decreases by 7\%. 

In all three cases studied here there is some effect of a (potential) companion on the derived luminosity and effective temperature from the
SED fitting. The effects are also systematic in nature. However, even for quite small contrast levels ($3.3 - 3.7$ mag between cepheid and companion)
the effects are (much) smaller than the random errors quoted on $L$ and $T_{\rm eff}$.
The effect of photometric contamination by a companion should have a small to negligible influence on the results in this paper.

\subsection{A comparison of effective temperature and reddening values to the literature}
\label{S-BIN}

The effective temperatures in the present work are derived by fitting model atmospheres to the SEDs (constructed to be representative of mean light),
which are dereddened taking reddening values from the literature.

Effective temperature have been derived from spectroscopy for many stars in the sample. The case of the five cepheids in
the inner disk (Sect.~\ref{S-IND}) illustrated the sensitivity of the photometric temperature determination on the reddening.
For both parameters it is interesting to compare the adopted reddenings and the derived effective temperatures to independently determined values.

In the case of the effective temperatures the results of \cite{Luck18} are used, which is by far the largest collection of uniformly reduced and analysed
spectra for CCs, including multi-epoch data when available. Table~3 from that paper is used, and for the 432 stars in overlap with the present
sample the following quantities are determined: number of epochs, and the minimum, maximum, average and median effective temperature.
Inspecting the results for the cepheids with the most multi-epoch data (also see Figs.~11-17 in \citealt{Luck18}) indicate
that the highest effective temperatures are found
in the phase range 0.9-0.1, and the lowest in the range 0.4-0.6. As the temperature from the SED fitting should be representative of mean light the
number of epochs and the average temperature in the phase range 0.1-0.4 and 0.6-0.9 is also calculated.
The same procedure is followed for ID~1-5 using the data in I19. 

\begin{figure}
\centering

\begin{minipage}{0.44\textwidth}
\resizebox{\hsize}{!}{\includegraphics{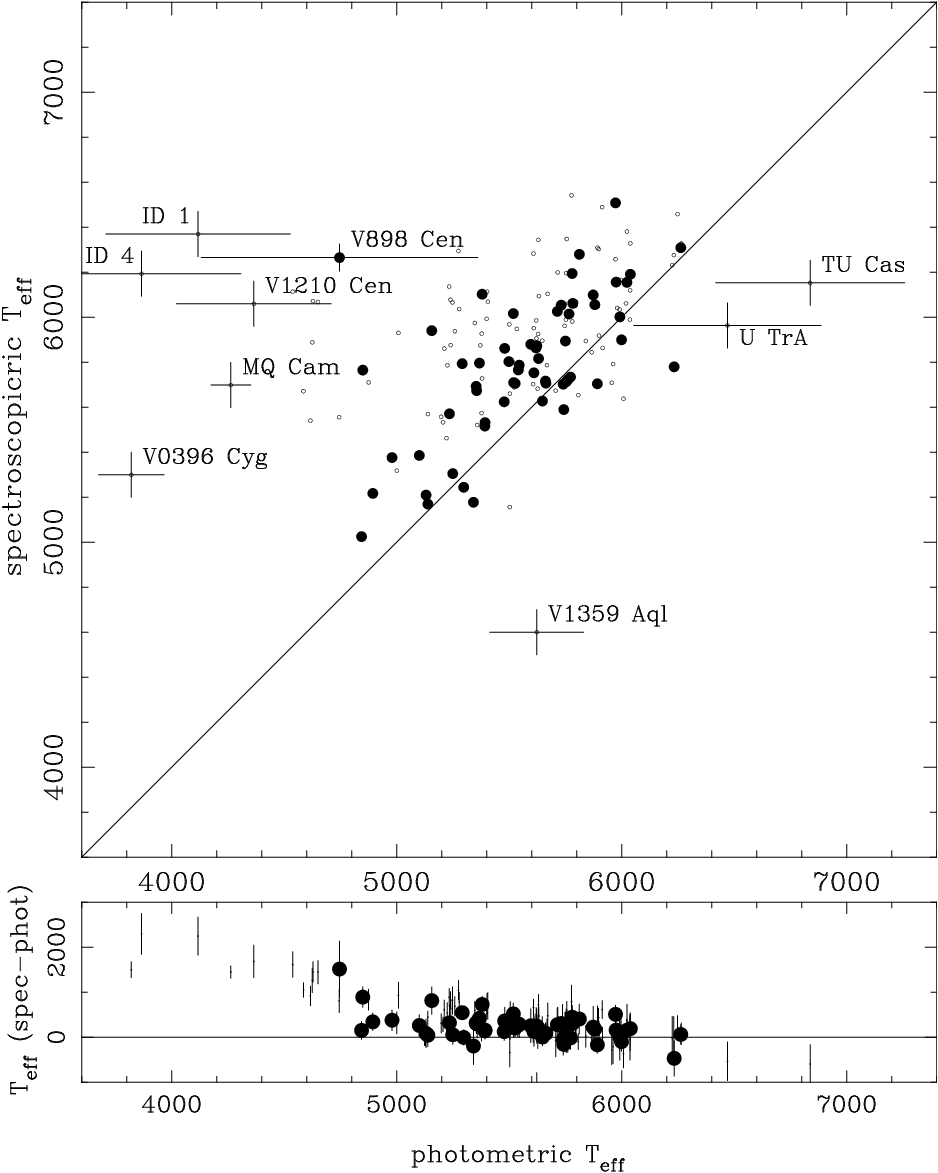}}
\resizebox{\hsize}{!}{\includegraphics{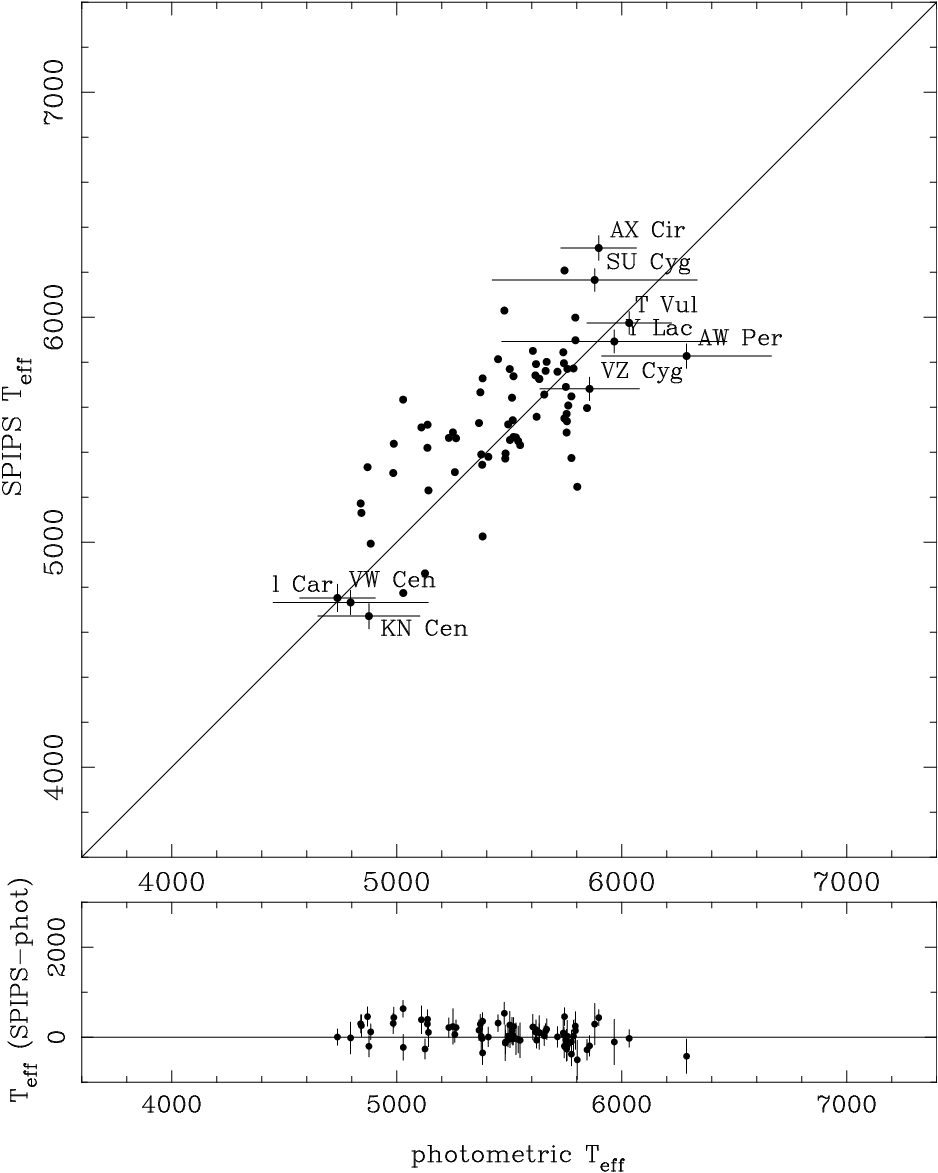}}
\end{minipage}

\caption{Comparison between the effective temperatures determined from the SED fitting, and in the literature.
Stars outside the bulk of objects are identified and plotted with error bars. The one-to-one line is indicated.
%
In the upper figure the temperatures are compared to the data in \cite{Luck18} (and I19 for ID~1-5).
When there are three or more values available in the phase range representative of mean light the object is marked
by a thick filled circle. Otherwise when there are two or more observations the average is taken (small open circles).
In the lower panel the difference between spectroscopic and photometric temperature is plotted.
In the lower figure the effective temperatures from \citet{Trahin19} are compared to the present work.
}
\label{Fig:Teff}
\end{figure}

In the upper panel of Fig.~\ref{Fig:Teff}, the spectroscopic temperatures are compared to the photometric ones.
If there are more than three determinations in the phase range representative of mean light the average over those values is
taken (the thick solid circles),
otherwise when there are two or more observations the average is taken (small open circles).

Some interesting features can be observed. The best determined spectroscopic temperatures correlate with the photometric
determinations, but there is an offset of 200 $\pm$ 235~K (58 stars, excluding V898 Cep), and the panel with the residuals
even suggests a trend, for which no explanation is apparent.

When the stars with two or more determinations are considered as well the scatter increases (as expected), but some very clear outliers also
appear. All five stars that are significantly hotter than expected have exactly two measurements, and since the temperature changes over the
pulsation cycle this could be a statistical effect.

\begin{figure}
\centering

\begin{minipage}{0.49\textwidth}
\resizebox{\hsize}{!}{\includegraphics{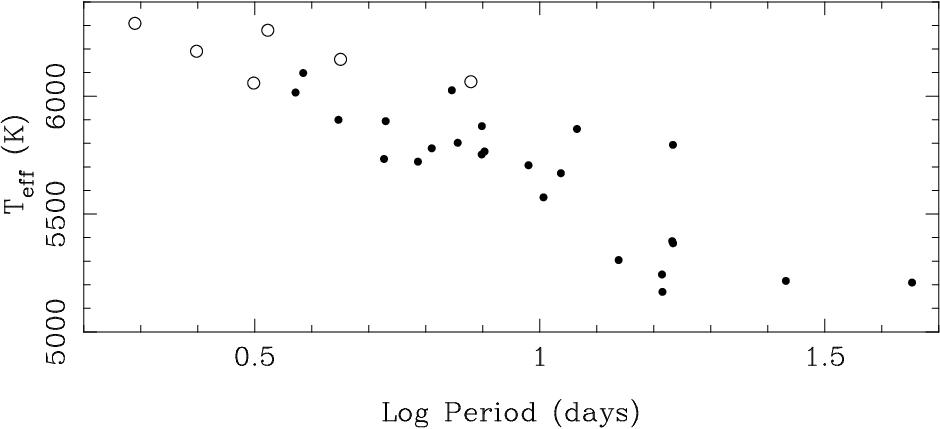}}
\resizebox{\hsize}{!}{\includegraphics{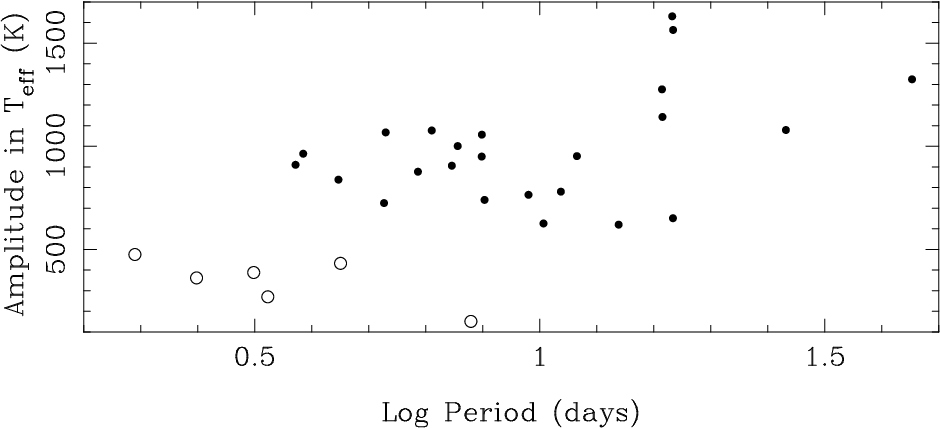}}
\resizebox{\hsize}{!}{\includegraphics{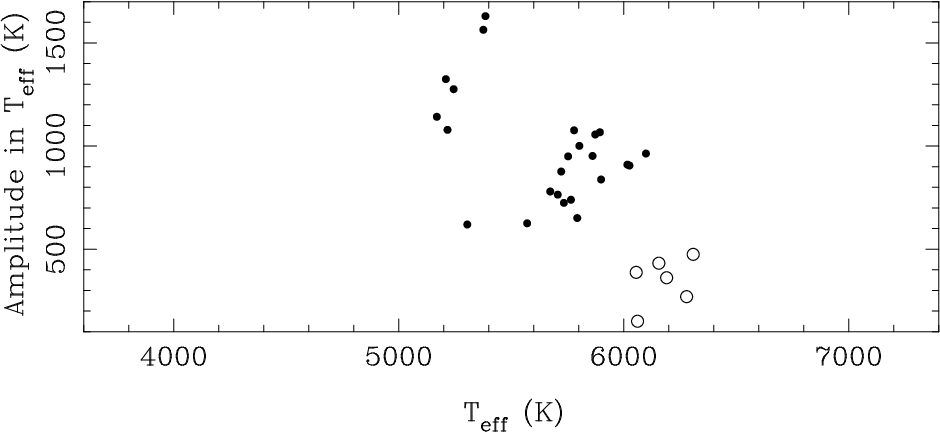}}

\end{minipage}

\caption{
  Using data from \cite{Luck18} the panels show the mean effective temperature, and the difference in $T_{\rm eff}$ over
  the pulsation cycle against temperature and period for the 30 objects with seven or more datapoints in the phase range
  typical for mean light. Overtone pulsators are indicated by open circles.
}
\label{Fig:TeffAmpl}
\end{figure}

Figure~\ref{Fig:TeffAmpl} shows the mean spectroscopic effective temperature versus period, and the difference between highest and lowest
effective temperature over the pulsation cycle versus period and $T_{\rm eff}$. The data is from  \cite{Luck18} and considers the 30 stars
with seven or more determinations in the phase range typical of mean light. 
The top panel is quasi identical to that of Fig.~18 in \cite{Luck18}. He used 52 stars (his criterion was five or more determinations in total) but did
not distinguish between FU and FO pulsators.
The two other panels show the range in effective temperature over the pulsation cycle as a function of temperature and period.
Overtone pulsators show changes that are about a factor of three smaller than FU pulsators at the same period.
The plot shows that changes of 1000~K over the pulsation cycle are quite common and possibly even higher at lower effective temperatures.

This indicates that the location of the five outliers can only partly be explained by the sampling of the two datapoints over the light curve.
Another indication that the photometric temperature might be incorrect is that all these five stars
(and to a lesser extent TU Cas and U TrA) are outliers in the HRD as well and that the spectroscopic temperature would put all these objects
closer to the IS. Given the discussion in Sect.~\ref{S-IND} one might reddening to play a role; this is investigated in the next section.

Another comparison of the effective temperatures is with the recent work of \citet{Trahin19}, who applied the
Spectro-Photo-Interferometry of Pulsating Stars (SPIPS) method \citep{Merand2015} to a sample of 74 CCs (and that all are in the present sample).
In the SPIPS method light curves in different bands, radial velocity curves, spectroscopic temperature determinations, and angular diameter determinations
are fitted to provide a consistent model fit to all data. 
What is interesting in the present context is that effective temperature (via ATLAS9 model atmospheres) and reddening are fitted simultaneously.
The bottom panel in Fig.~\ref{Fig:Teff} shows the comparison between the effective temperatures found here and in \citet{Trahin19}.
The agreement is very good. The offset of $66 \pm 230$~K is not significant.
The scatter suggests that the error bars in \citet{Trahin19} may be underestimated as the median error bar among the 74 stars is 52~K while
  it is 188~K in the temperature determinations derived in the present work.

Although different in detail, both SPIPS and the present work use grids of (different) model atmospheres to fit photometry.
That the effective temperatures agree to within the errors with no significant offset is highly satisfactory.

\subsection{Reddening}
\label{S-EBV}

\begin{figure}
\centering

\begin{minipage}{0.49\textwidth}
\resizebox{\hsize}{!}{\includegraphics{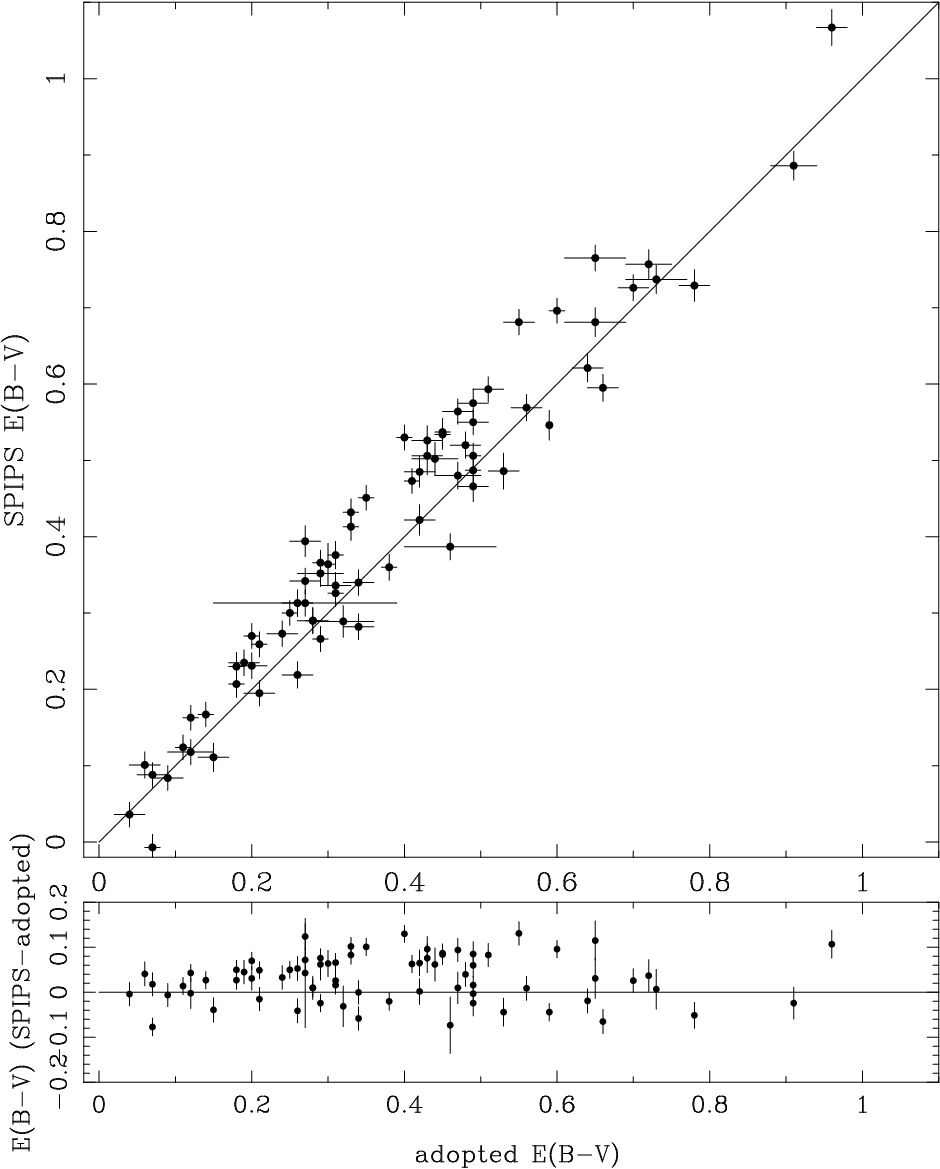}}
\end{minipage}

\caption{Comparison between the $E(B-V)$ values in \citet{Trahin19} and the present work.
The one-to-one line is indicated.  The random and systematic error bars in \citet{Trahin19} have been added in quadrature.
}
\label{Fig:EBV}
\end{figure}

The discussion on the cepheids in the direction of the inner disk and the discrepancy in some cases between spectroscopic
and photometric temperature determinations suggests that reddening could play a role.
Figure~\ref{Fig:EBV} compares the $E(B-V)$ values determined in \citet{Trahin19} to the values adopted from the literature
in the present work. The results are overall consistent with no significant outliers.
A linear least-squares fit gives a slope not significantly different from unity: 
$E(B-V)_{\rm SPIPS} = (1.02 \pm 0.03)\; E(B-V)_{\rm this\;work} +  (0.027 \pm 0.012)$, with an rms of 0.051.
This scatter is larger than might be expected based on the error bars in the two measurements. The median error bar in the reddening
  in these 74 stars is 0.02~mag in this work and 0.017~mag in \citet{Trahin19}, suggesting that both errors are on average underestimated.

However, none of the two dozen outliers marked in Fig.~\ref{Fig:Teff} are in the sample of \citet{Trahin19} and
therefor the role of reddening can not be excluded for these specific stars.

To investigate the matter further two recent 3D reddening models have been used to estimate the reddening in the direction of the stars in the sample.

The first is described in \cite{Lallement18}\footnote{\url{https://stilism.obspm.fr/} (version 4.1).
} (hereafter STILISM) and is based on {\it Gaia}, 2MASS and APOGEE-DR14 data.
For a given galactic longitude, latitude and distance, the tool returns the value of $E(B-V)$ and error, and the distance to which these values refer.
If this distance is smaller than the input distance the returned value for the reddening is a lower limit.
In these cases a simple estimate of the reddening at the distance of the cepheid is made.
A second reddening value is queried at a distance 0.75 times the maximum distance available in the grid in that direction.
Based on this the first derivative (with error bar) is determined and the reddening at the distance of the target estimated.
The error bar returned by STILISM is added in quadrature with the error due to a 1 degree change in $l$ and $b$, and a 5\% error in distance.

The second reddening model is that described in \citet{Green2019}\footnote{\url{https://argonaut.skymaps.info} The `Bayestar19' dataset.}
and is based on {\it Gaia} DR2 data, 2MASS, and Pan-STARRS~1 data.
Reddening values are provided out to `several' kpc for stars north of declination $-30\degr$, which means 340 stars in the sample.
The output requested from the code are the 2.1, 50 and 97.9\% probability values of the reddening. The two extremes are used to calculate the error in the
reddening value. This error is added in quadrature to the error due to a 1 degree change in R.A. and Declination, and a 5\% error in distance.

The adopted $E(B-V)$ values from the literature, and those from \citet{Trahin19} and the two reddening models, are collected in Table~\ref{Tab:App}.
The different sets are briefly compared and discussed in Appendix~\ref{AppRedd}.
For some of the outliers in Figs.~\ref{Fig:HRD} and \ref{Fig:Teff} the alternative reddening values are very different from the adopted ones,
but not in all cases.

For two of the most prominent outliers in Fig.~\ref{Fig:HRD}, DV Ser and V891 Per,
the two 3D reddening models agree and give a
$E(B-V)$ value very different from the adopted ones. Redoing the fitting for an $E(B-V) = 1.4$ results in luminosities and effective temperatures
that puts both stars inside the IS.
However there are also outliers where the two reddening models and the adopted value agree (e.g. TU Cas, ASAS 1810-20), or
where the two reddening models do not agree among them, but one of them agrees with the adopted reddening (e.g. BD Cas, IR Cep), or
where the adopted value agrees with the single available value from a reddening model (e.g. V1210 Cen).

\subsection{Dust and excess emission}
\label{S-EXC}

Some stars in the sample have been proposed to show infrared excess which was suggested to be due to dust emission in a
circumstellar envelope (CSE), see e.g. \citet{Merand2005,Merand2006,Kervella2006,Merand2007,Barmby11,GalVISIR,Gallenne13b,Merand2015,Breitfelder16}.

In this section, the results are presented of a consistent model of the cepheid surrounded by a (spherically symmetric) dust shell
using MoD. An important ingredient to such a model is the dust opacity. Other parameters are the dust optical depth ($\tau$),
the inner radius of the CSE (or the dust temperature at $R_{\rm inn}$), and the slope of the density law $\sim r^{-p}$.

\citet{Gallenne13b} also performed dust RT calculations to fit the SED and MIDI MIR spectra of
T Mon and X Sgr. These two stars seem to be the only CCs for which quantitative RT calculations have been performed so far.
They investigated combinations of several dust species and based on their results a similar approach was adopted
and grains composed of metallic iron (optical constants from \citealt{Henning95}), warm silicates  \citep{OH94}, and
compact aluminium oxide  \citep{Begemann97} are considered. A grain size of 0.1~$\mu$m is adopted and the absorption and
scattering coefficients are calculated assuming a distribution of hollow spheres \citep{Min05}.

\begin{figure}

\begin{minipage}{0.24\textwidth}
\resizebox{\hsize}{!}{\includegraphics{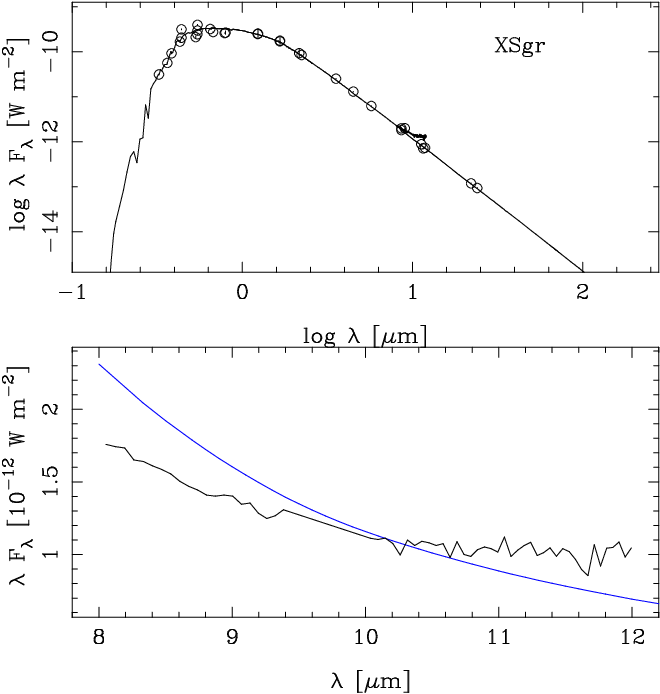}}
\end{minipage}
\begin{minipage}{0.24\textwidth}
\resizebox{\hsize}{!}{\includegraphics{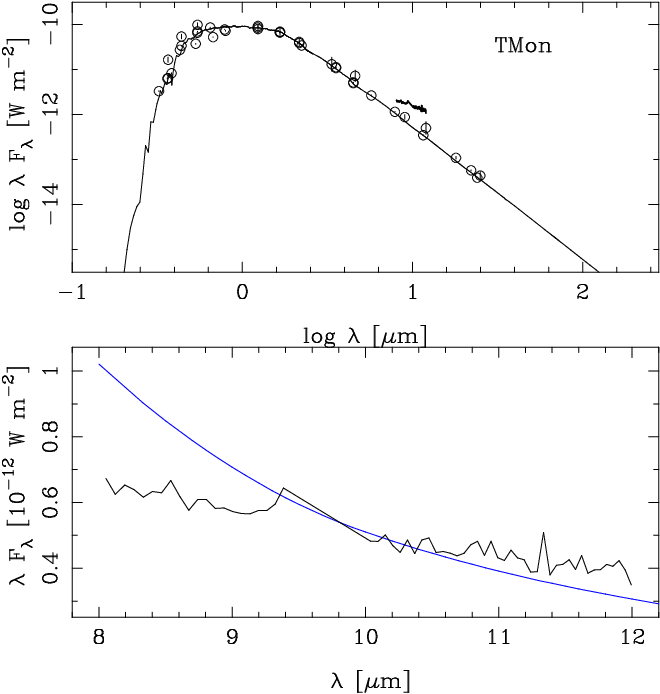}}
\end{minipage}

\begin{minipage}{0.24\textwidth}
\resizebox{\hsize}{!}{\includegraphics{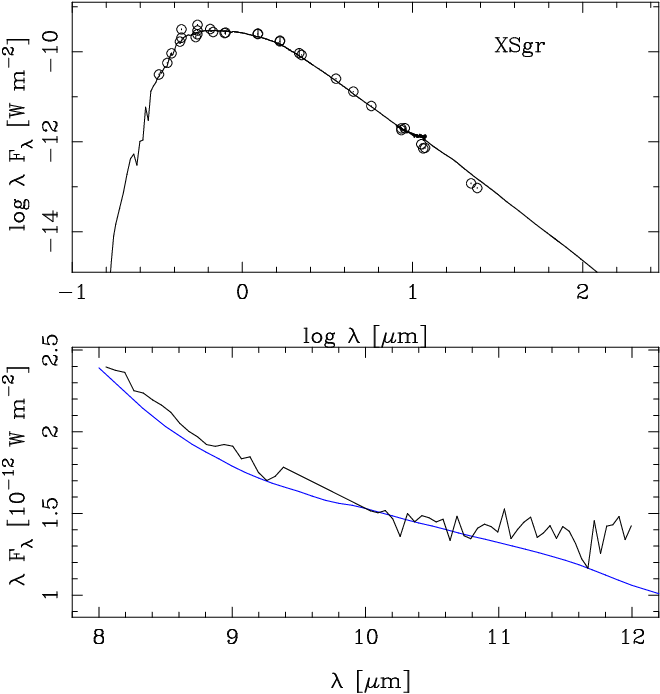}}
\end{minipage}
\begin{minipage}{0.24\textwidth}
\resizebox{\hsize}{!}{\includegraphics{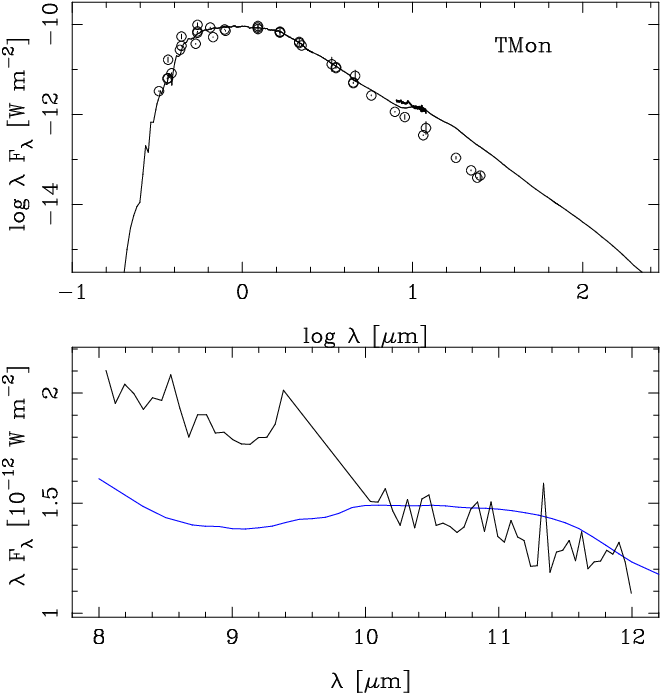}}
\end{minipage}

\begin{minipage}{0.24\textwidth}
\resizebox{\hsize}{!}{\includegraphics{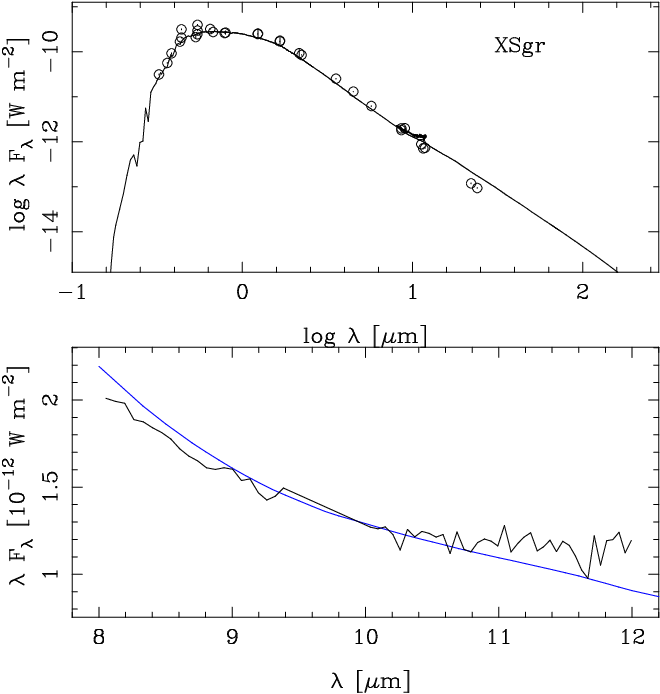}}
\end{minipage}
\begin{minipage}{0.24\textwidth}
\resizebox{\hsize}{!}{\includegraphics{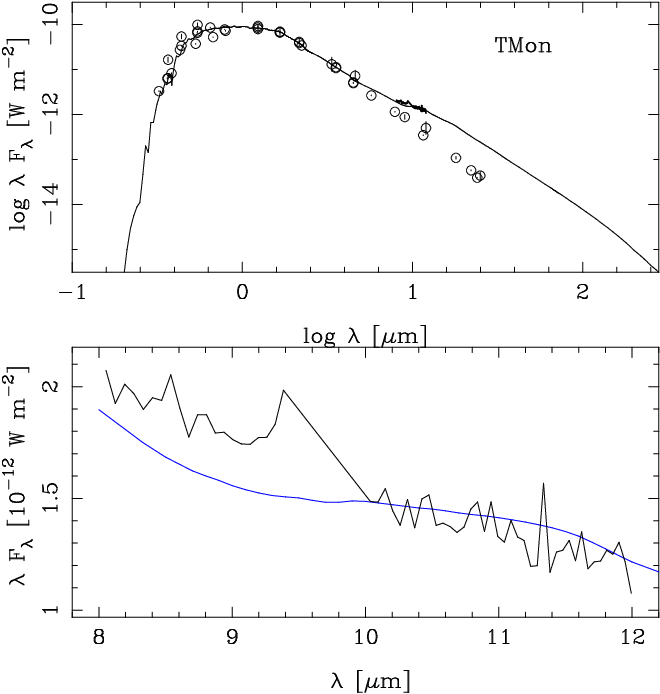}}
\end{minipage}

\begin{minipage}{0.24\textwidth}
\resizebox{\hsize}{!}{\includegraphics{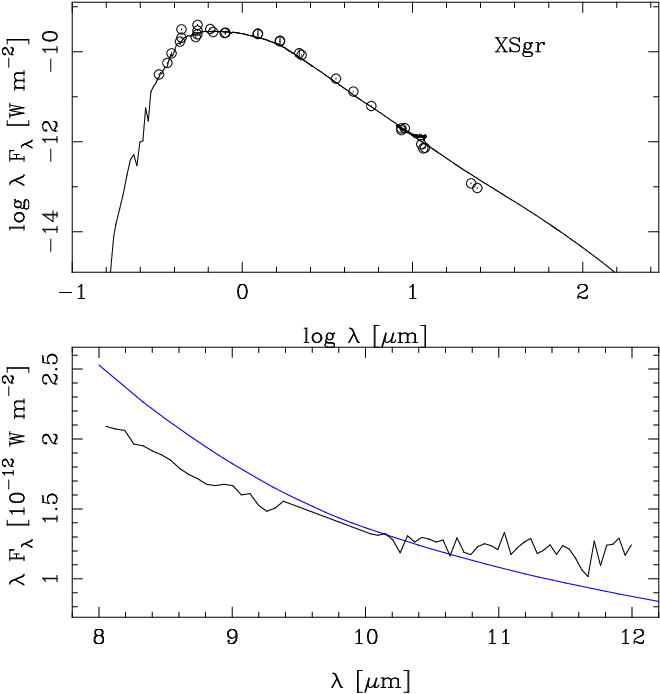}}
\end{minipage}
\begin{minipage}{0.24\textwidth}
\resizebox{\hsize}{!}{\includegraphics{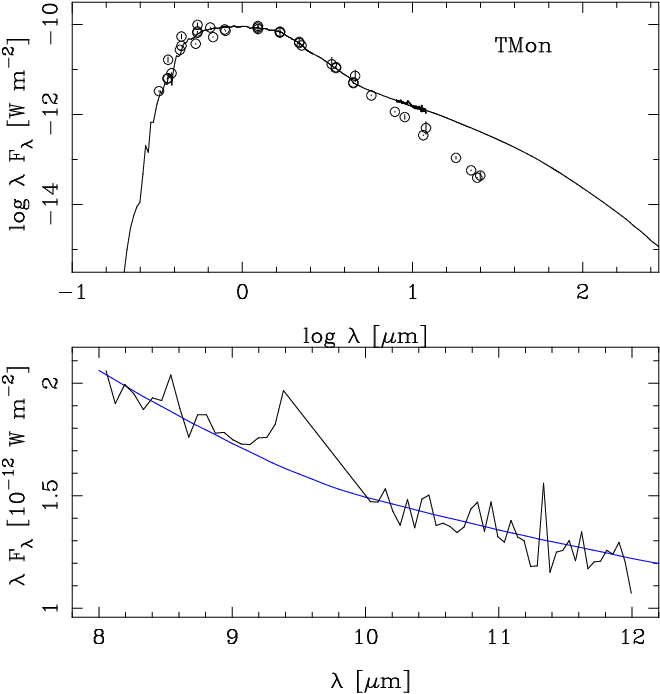}}
\end{minipage}

\caption{Fits to the SED and mid-IR spectra of T Mon and X Sgr for different dust compositions.
The top panel is the standard model without dust component.
The other three panels are models with dust composed of 80, 94, and 100\% iron, respectively, with the
remainder evenly split between silicates and aluminium oxide. 
The mid-IR spectrum is scaled to the model based on the average flux in the 10-10.5~$\mu$m region.
}
\label{Fig:TMXS}
\end{figure}

The results of the calculations are shown in Fig.~\ref{Fig:TMXS}.
The top panel shows the standard case, a model with no dust, and three models with dust composed of (top to bottom)
80, 94, and 100\% iron and the remainder evenly split between silicates and aluminium oxide.
One immediate notices that the MIDI spectrum of T Mon is not compatible with the rest of the SED.
The SEDs for T Mon and X Sgr in the present paper have more datapoints (also in the 10-20~$\mu$m region) than
considered in \citet{Gallenne13b}. However, also in \citet{Gallenne13b} the spectrum lies well above their SED extrapolated from
shorter wavelengths, and in fact, they caution that `the excess of T Mon \ldots 
might suffer from skybackground contamination'.
This is indeed likely to be the case. Although one can fit the shape and flux level of the MIDI spectrum with
featureless pure iron dust the fit to the photometric points excludes that the observed MIDI spectrum is associated to the star.
As this discrepancy was noted early on in this study, the spectrum of T Mon was down-weighted when performing
the standard fit without dust, not to influence the determination of effective temperature and luminosity.

The shape of the MIDI spectrum of X Sgr can be fit reasonably well with dust composed of 80 or 94\% iron, similar to the
results in \citet{Gallenne13b}. The temperature at the inner radius is found to be 1309 $\pm$ 40~K, corresponding
to a size of 18 mas, and the dust optical depth at 0.55~$\mu$m to be $(15 \pm 1) \; 10^{-3}$ also in agreement
with the values of, respectively, $1684 \pm 225$~K, $(7.9 \pm 2.1) \; 10^{-3}$, and $(15.6 \pm 2.9)$ mas in \citet{Gallenne13b}.

\begin{figure*}

\begin{minipage}{0.24\textwidth}
\resizebox{\hsize}{!}{\includegraphics{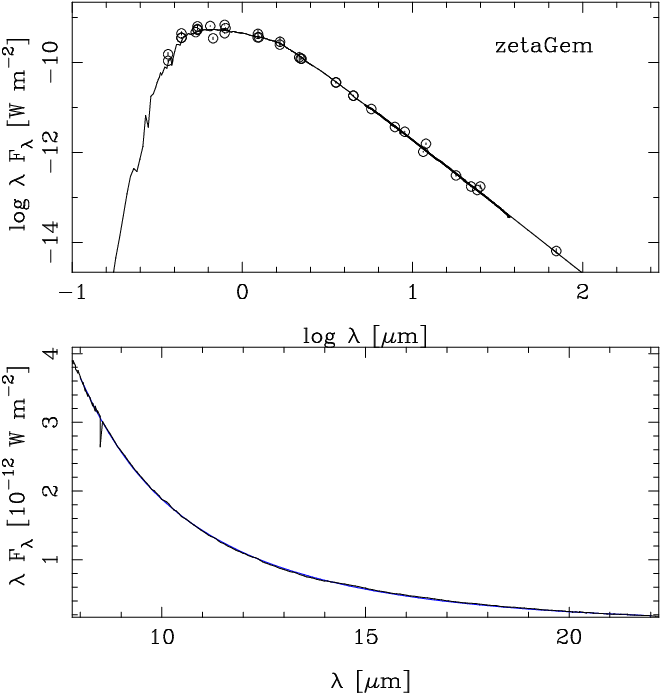}}
\end{minipage}
\begin{minipage}{0.24\textwidth}
\resizebox{\hsize}{!}{\includegraphics{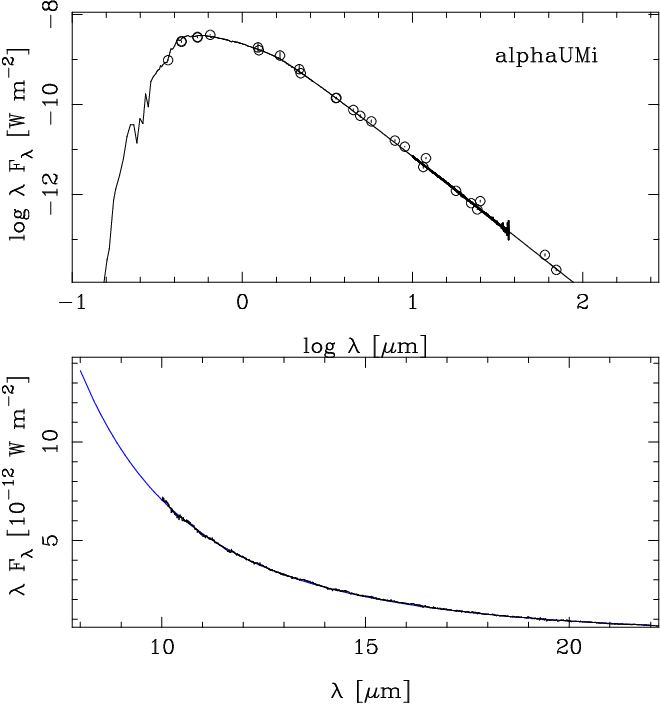}}
\end{minipage}
\begin{minipage}{0.24\textwidth}
\resizebox{\hsize}{!}{\includegraphics{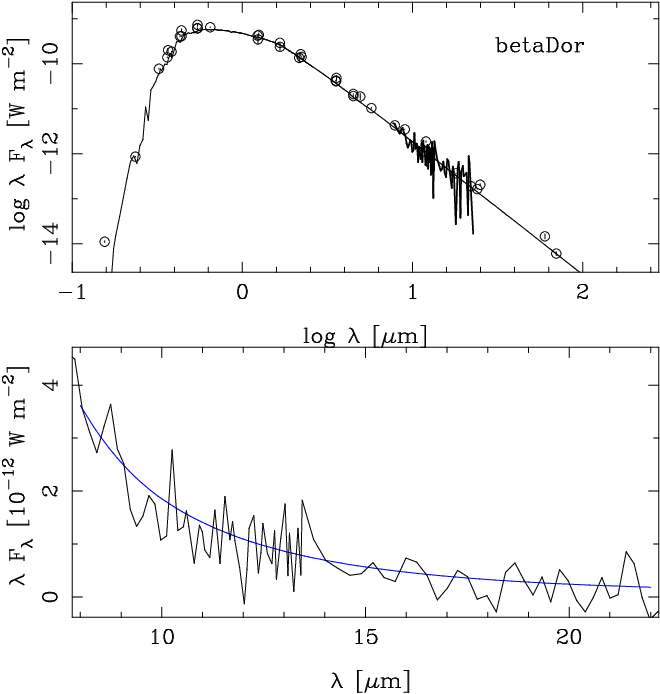}}
\end{minipage}
\begin{minipage}{0.24\textwidth}
\resizebox{\hsize}{!}{\includegraphics{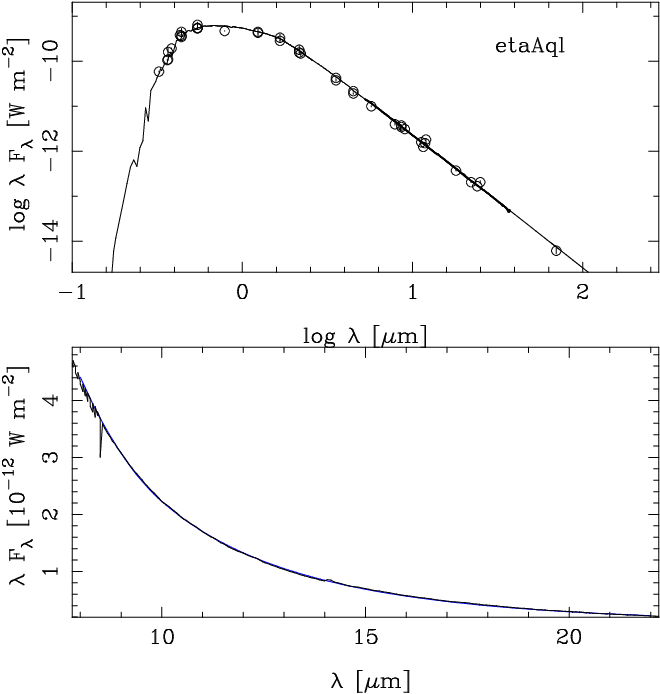}}
\end{minipage}

\begin{minipage}{0.24\textwidth}
\resizebox{\hsize}{!}{\includegraphics{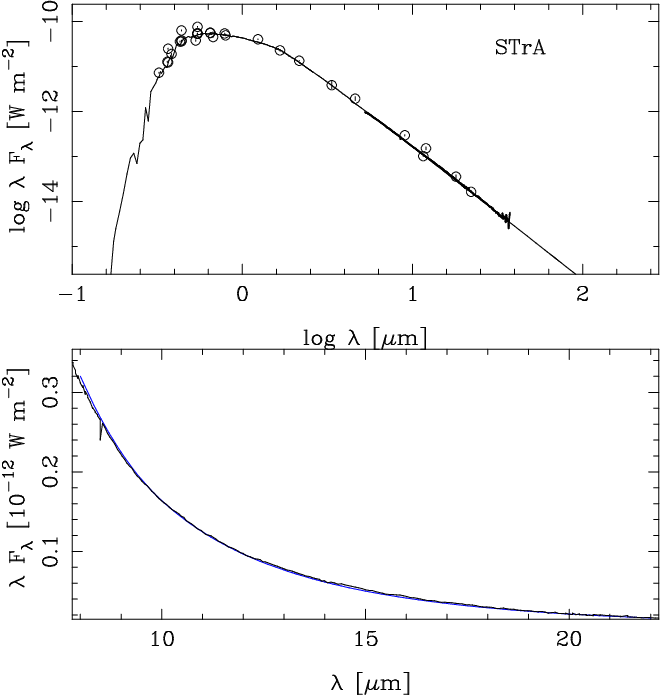}}
\end{minipage}
\begin{minipage}{0.24\textwidth}
\resizebox{\hsize}{!}{\includegraphics{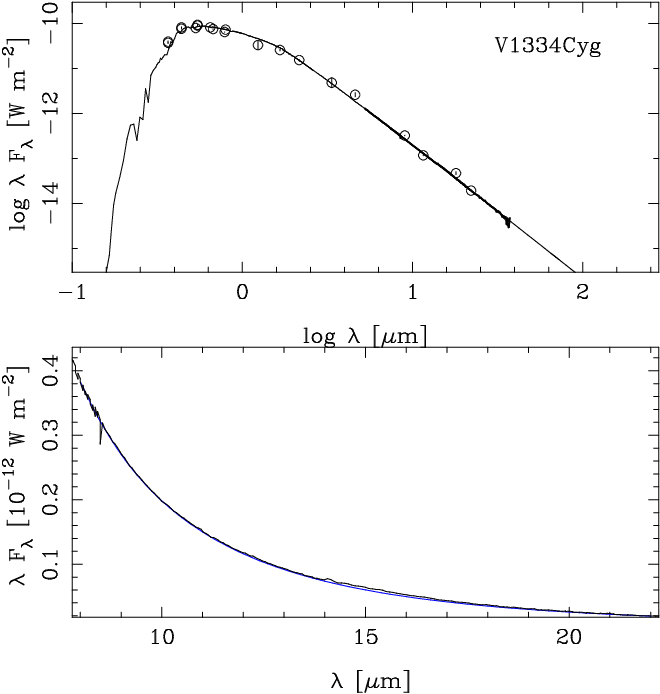}}
\end{minipage}
\begin{minipage}{0.24\textwidth}
\resizebox{\hsize}{!}{\includegraphics{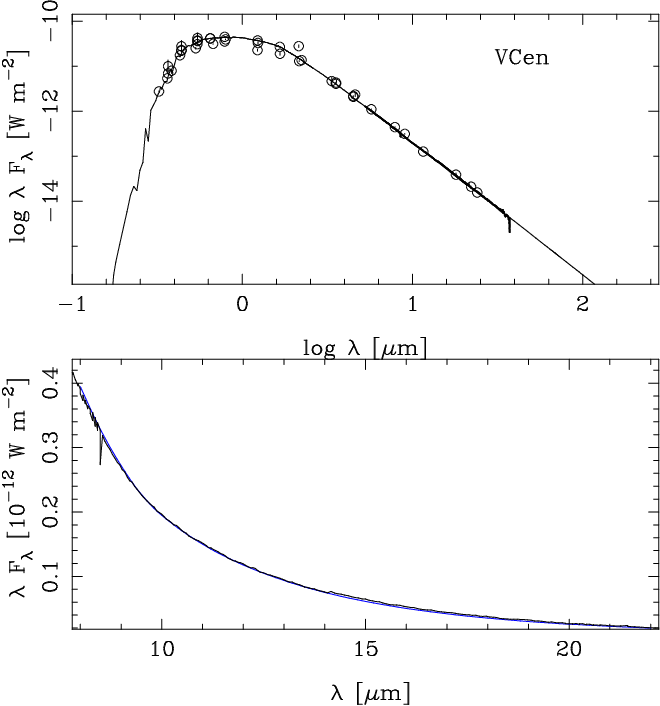}}
\end{minipage}
\begin{minipage}{0.24\textwidth}
\resizebox{\hsize}{!}{\includegraphics{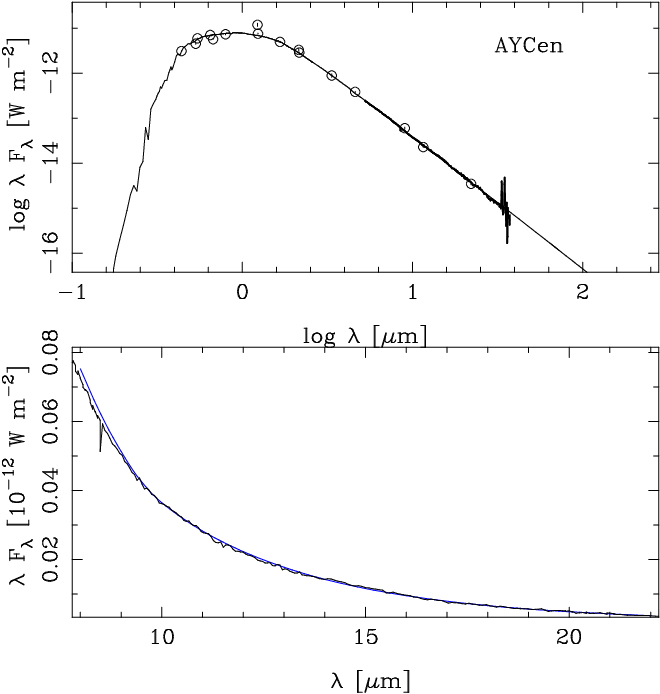}}
\end{minipage}

\begin{minipage}{0.24\textwidth}
\resizebox{\hsize}{!}{\includegraphics{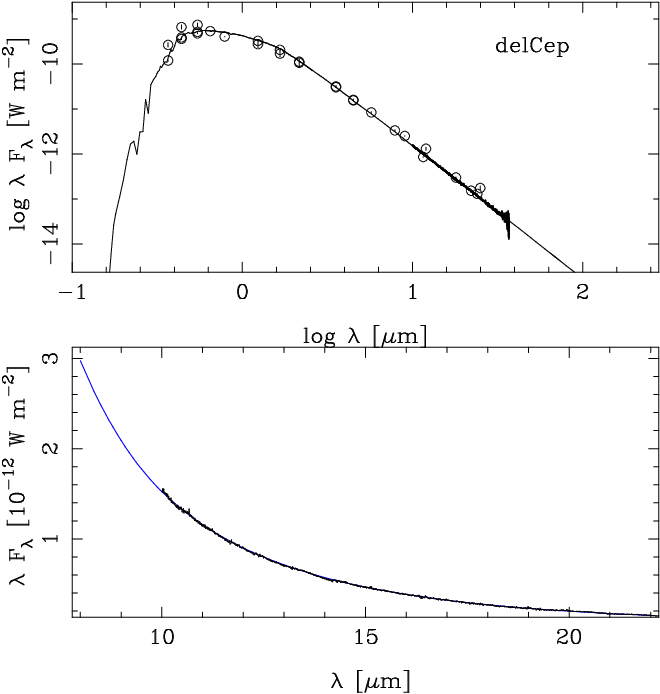}}
\end{minipage}
\begin{minipage}{0.24\textwidth}
\resizebox{\hsize}{!}{\includegraphics{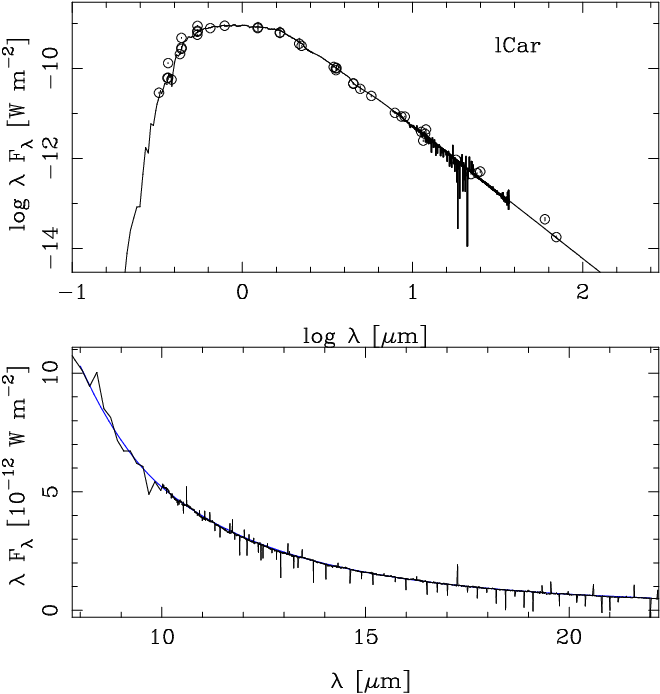}}
\end{minipage}
\begin{minipage}{0.24\textwidth}
\resizebox{\hsize}{!}{\includegraphics{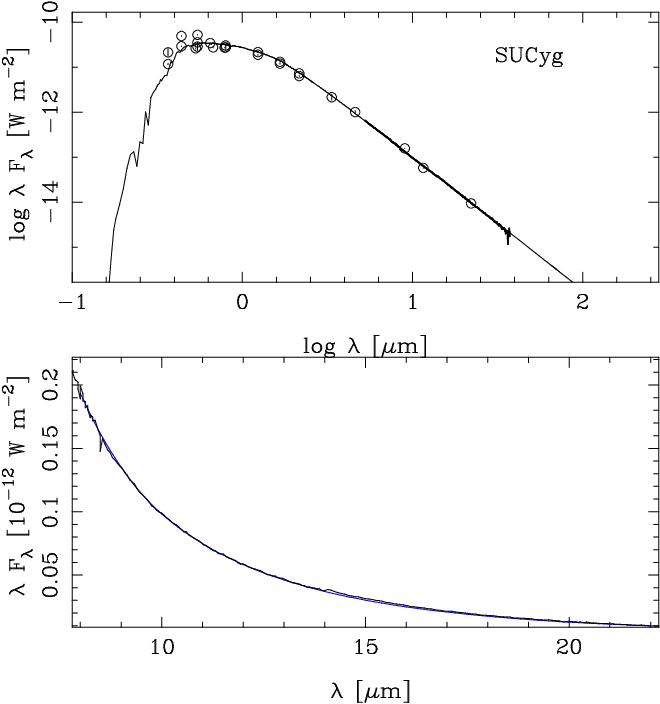}}
\end{minipage}
\begin{minipage}{0.24\textwidth}
\resizebox{\hsize}{!}{\includegraphics{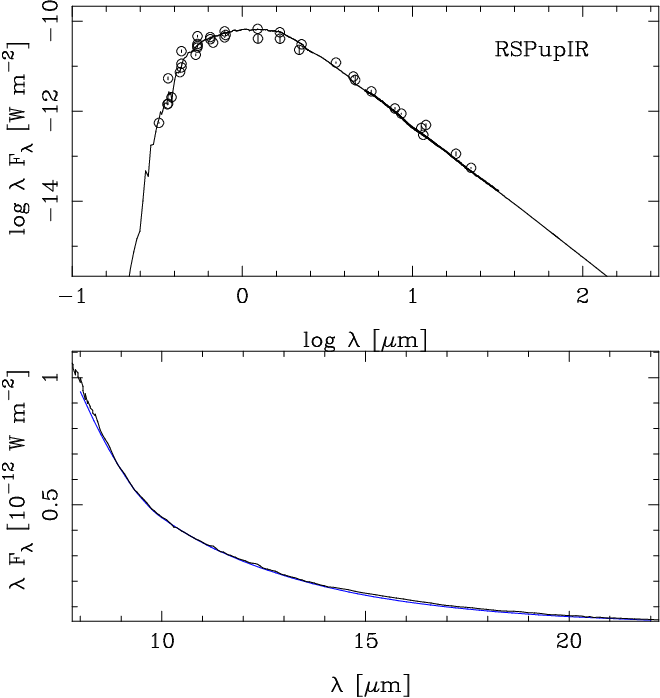}}
\end{minipage}

\caption{Fits to the SED and mid-IR spectra of the cepheids which have an {\it IRAS} LRS or {\it Spitzer} IRS spectrum.
The mid-IR spectrum is scaled to the model based on the average flux in the 10-10.5~$\mu$m region.
}
\label{Fig:IRS}
\end{figure*}

Based on these results the other CCs with MIR spectra are fitted with dust composed of 
80, 90, 94, and 100\% iron dust. The temperature at the inner radius and the slope of the density law
were kept as free parameters, unless no convergence was achieved and $p$ or $T_{\rm inn}$, or both, were fixed.
The results are collected in Table~\ref{Tab:IR} and the best fits are shown in Fig.~\ref{Fig:IRS}.
The table first lists the luminosity and the statistics ($\chi_{\rm r}^2$ and BIC) for the model without dust and then
the parameters for the best-fitting model with dust. In the case of V1334 Cyg a model was also run on the photometry
corrected for the binary component (Sect.~\ref{S-BIN}).
The error bars on the luminosities are much smaller for the same star than those listed in Table~\ref{Tab-Targets}.
  The reason is that the error in Table~\ref{Tab-Targets} includes the error in the effective temperature, while the error in Table~\ref{Tab:IR} is
  that when the effective temperature is fixed to its best-fitting value.
  The errors on the luminosities are very small. The reason is the much larger number of available data points compared to the stars without spectral information.
  There are typical $\sim 380$ datapoints contained in an IRS spectrum and $\sim 18-30$ photometric datapoints.
  With a typical residual of 0.1~mag per data point one can estimate an error on the mean of order 0.5\%. 
In three stars, no converging model or no significant improvement in the fit is obtained; in the other cases, a statistically
better fit can be obtained by including a dust component.

One very interesting observation is that the best-fitting luminosity in the dust model is lower than in the standard case and 
that (in most cases) the $V$ and $K$ magnitudes in the model with dust are fainter than the model without dust.
The excess compared to the photosphere is relatively small in the models with dust, 10~mmag at most in $K$.
Its larger in the $N$-band, 10-40~mmag but smaller than the few percent claimed in the literature for both bands.

Within the assumptions of the adopted dust model one can relate the optical depth to a mass-loss rate.
Assuming a dust-to-gas ratio of 1/200 and an expansion velocity of the CSE of 200~\ks\
(the escape velocity for a 5~\msol\ 45~\rsol\ star) the mass-loss rates is about $3.6 \cdot 10^{-9}$ \msolyr\
in the case of AY Cen, and factors of 10-100 lower in the other stars.

\begin{sidewaystable*}
  \small
\setlength{\tabcolsep}{1.2mm}
\caption{Infrared excess in CCs with mid-IR spectra.}
\begin{tabular}{rrrrrrrrrrrrrrrrrrl} \hline \hline 
  Name    &    $L$          & $\chi_{\rm r}^2$ & BIC &         $L$       &    $\tau$        &  $T_{\rm inn}$ & $R_{\rm inn}$ &     $p$      & Gr. & $\chi_{\rm r}^2$ & BIC & $\Delta M_{\rm bol}$ & $\Delta V$ & $\Delta K$ &  $\Delta V$ & $\Delta K$ & $\Delta N$ & Remarks   \\
          &    (\lsol)      &                      &     &       (\lsol)     & ($\cdot 10^{-4}$) &       (K)     &  (\rstar)  &              & (\%) &          &    &    (mag)   &     (mag)   &    (mag)  &         \\
\hline 
$\zeta$ Gem  & 3204 $\pm$ 2.3  &  38.7 &  -2183 &  3168 $\pm$  4.7 & 13.7 $\pm$ 1.8 & 1546 $\pm$ 187 & 13 & 1.90 $\pm$ 0.06 & 94 & 32.5 &  -4847 & -0.012 & -0.013 & -0.010 &  0.000 & -0.003 & -0.013 & \\ 
$\eta$ Aql   & 3008 $\pm$ 2.0  &  24.0 &  -8050 &  2970 $\pm$  3.9 &  9.0 $\pm$ 0.8 & 1394 $\pm$  98 & 11 & 2.02 $\pm$ 0.05 & 90 & 19.5 & -10028 & -0.014 & -0.015 & -0.013 &  0.001 & -0.002 & -0.013 & \\ 
$\alpha$ UMi & 2413 $\pm$ 1.2  &  32.6 & -19409 &  2260 $\pm$ 10.1 & 14.1 $\pm$ 1.1 & 1091 $\pm$  37 & 50 & 2.58 $\pm$ 0.03 & 80 & 26.8 & -27673 & -0.071 & -0.072 & -0.067 &  0.001 & -0.004 & -0.046  & \\ 
$\beta$ Dor  & 2863 $\pm$ 89   & 243.6  &  24737 &                  &                &                &     &            &    &       &        &  &       &         &       & & & no convergence \\ 
$\delta$ Cep & 1984 $\pm$ 0.90 &  16.8 & -45576 &  1955 $\pm$  6.0 &  6.2 $\pm$ 0.4 & 1218 $\pm$  69 & 32 & 2.26 $\pm$ 0.03 & 90 & 16.7 & -45726 & -0.016 & -0.017 & -0.015 &  0.000 & -0.001 & -0.012 &  \\ 
$l$ Car    & 12316 $\pm$ 7.9  &  31.2 & -22071 & 12313 $\pm$  8.4 &  1.6 $\pm$ 0.9 & 2825 $\pm$ 938 & 1.8 & 2.0        f    & 80 & 31.2 & -22144 & -0.000 &  0.000 &  0.000 &  - & - & - & insignificant \\ 
AY Cen    & 1864 $\pm$ 3.2  &  13.1 & -15069 &  1808 $\pm$  6.0 & 48.6 $\pm$ 8.6 & 1590 $\pm$ 154 & 14 & 2.39 $\pm$ 0.13 & 80 &  9.4 & -16561 & -0.033 & -0.035 & -0.023 &  0.002 & -0.010 & -0.040 &\\ 
SU Cyg    &  910 $\pm$ 0.90 &  72.7 &   8981 &   897 $\pm$  1.3 &  3.5 $\pm$ 0.6 & 1126 $\pm$ 115 & 44 & 2.10 $\pm$ 0.10 & 80 & 54.3 &   1257 & -0.016 & -0.014 & -0.013 &  0.000 & -0.001 & -0.014 &  \\ 
S TrA     &  486 $\pm$ 0.90 &  36.2 &  -4498 &   473 $\pm$  1.0 &  2.7 $\pm$ 0.4 &  512 $\pm$  36 & 370 & 2.88 $\pm$ 0.14 & 80 & 16.2 & -12352 & -0.029 & -0.029 & -0.029 &  0.000 & 0.000 & -0.038 & \\ 
V Cen     & 1657 $\pm$ 2.2  &  22.1 & -10162 &  1615 $\pm$  2.6 & 15.1 $\pm$ 1.4 & 1668 $\pm$ 122 & 11 & 1.77 $\pm$ 0.04 & 80 &  8.7 & -15908 & -0.028 & -0.029 & -0.025 &  0.001 & -0.002 & -0.029 & \\ 
V1334 Cyg & 1871 $\pm$ 1.7  &  25.2 &  -9630 &  1845 $\pm$  2.2 & 26.1 $\pm$ 2.5 & 3236 $\pm$ 382 & 2.1 & 1.76 $\pm$ 0.05 & 80 & 15.6 & -13564 & -0.015 & +0.054 & -0.012 &  & & & \\ 
          & 1764 $\pm$ 1.6  &  24.5 &  -9928 &  1740 $\pm$  2.1 &  2.8 $\pm$ 0.2 &  979 $\pm$  48 & 63 & 1.96 $\pm$ 0.06 & 80 & 14.4 & -13915 & -0.015 & -0.015 & -0.015 & -0.002 & -0.003 & -0.017  & binary corrected \\ 
RS Pup   & 13778 $\pm$ 14.5 &  36.2 &  -1805 & 13727 $\pm$ 16.6 &  5.4 $\pm$ 2.5 & 1300 f         & 18 & 2.0 f           & 90 & 36.2 & -1813  & -0.004 & -0.004 & -0.004 &  - & - & - & insignificant \\ 
\hline
\end{tabular} 
\label{Tab:IR}
\tablefoot{
Column~1 gives the name of the cepheid.
Columns~2-4 give the luminosity (for the best-fitting effective temperatue), the reduced $\chi^2$ and the BIC statistics.
Columns~5-10 give the results for the best-fitting model including a dust component: luminosity, dust optical depth at 0.55~$\mu$m, temperature of the dust at the inner radius,
the corresponding inner radius in stellar radii, and
slope of the density law. An `f' in Cols.~7-9 means that parameter is fixed. The number in  Col.~10 indicates the percentage of iron in the grain (see main text), while
Cols.~11-12 give the statistics for this model.
Columns~12-14 gives the difference in magnitude between the standard case and the model with dust bolometrically, and in the $V$ and $K$. A negative magnitude implies that the model with dust is fainter.
Columns~15-17 gives the difference in magnitude between the model with dust and the photosphere in the $V, K$ and $N$ band. A positive magnitude implies that photosphere is brighter.
}  
\end{sidewaystable*}

Finally, the SEDs of all stars in the sample (except the three stars with a clear IR excess, FQ Lac, AU Peg, and QQ Per) are fitted with a dust model.
As there are, in general, no spectra available to better constrain the fitting, the slope of the density law is fixed to 2.0 to reduce the number of free parameters,
a value consistent with the results found for the objects with MIR spectra. 
Grains with 80\% iron (and thus 10\% silicate and 10\% aluminium oxide grains) are assumed.
Initial guesses for the optical depth are $\tau= 0.0002, 0.0007, 0.002$, and $T_{\rm inn}= 1000, 1500$~K are used based on the results in Table~\ref{Tab:IR}.
The models are run over the grid of model atmospheres for each of the six initial guesses of the dust parameters, and the model with the
lowest BIC is taken. It is compared to the BIC of the model without dust and the reduction in the BIC is compared to the change in BIC due
to a 1$\sigma$ change in effective temperature as a measure of the significance of the reduction in the BIC due to dust compared to other parameters.

\begin{figure}
\centering

\begin{minipage}{0.49\textwidth}
\resizebox{\hsize}{!}{\includegraphics{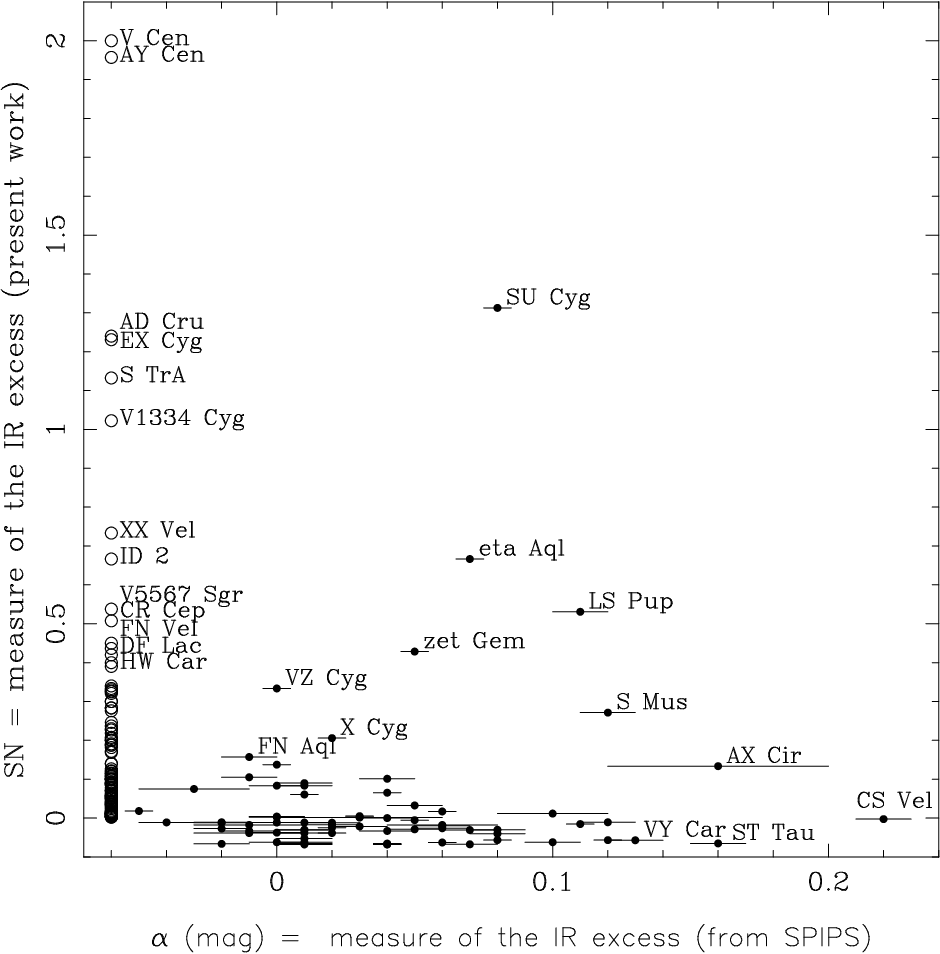}}
\end{minipage}

\caption{Measures of the IR excess from the present work, and a comparison to the measure of the IR excess from \citet{Trahin19}
(see main text for details).
}
\label{Fig:IR}
\end{figure}

For 331 stars in the sample, the models with dust do not converge or do not have a lower BIC.
The results are displayed in Fig.~\ref{Fig:IR}.
That figure also compares the results to the outcome of the SPIPS modelling from \citet{Trahin19}.
In that model an IR excess is parameterised using the functional form:
IRex= 0~mag for $\lambda < 1.2~\mu$m and IRex= $\alpha \cdot (\lambda - 1.2)^{0.4}$~mag  for $\lambda \ge 1.2~\mu$m,
where $\alpha$ is one of the outputs of the SPIPS model, and the quantity plotted along the abscissa in Fig.~\ref{Fig:IR}.
The value of $\alpha$ is effectively the excess in the $K$-band, while the excess in the $N$-band is about 2.5 times larger. 
Along the ordinate the ratio of the reduction in the BIC in the dust model divided by the reduction in BIC due to a 1$\sigma$
change in effective temperature is used (hereafter SN).
The stars in the sample that are not in \citet{Trahin19} are plotted at  $x= -0.06$.

There is a group of stars ($\zeta$ Gem, LS Pup, $\eta$ Aql, SU Cyg) for which there is a good correspondence between the two works
($\alpha$ $\more$  0.05~mag, and SN $\more$ 0.4). On the other hand, there are stars that have a large values for $\alpha$ for which
the SEDs are well defined in the present work and for which there is no evidence for IR excess (in particular CS Vel).

It is remarked that a non-negligible number of stars in \citet{Trahin19} are quoted to have a negative IR excess, which is physically impossible.
This is likely a testimony to the fact that it is very difficult to establish small levels of IR excess with confidence.
It may point to uncertainties in other aspects of the SPIPS modelling, for example the value of the $p$-factor.
It does suggest that the error bars quoted in \citet{Trahin19} for the IR excess are probably somewhat underestimated, and that the
range $-0.05 < \alpha\ \less +0.05$~mag is likely consistent with the absence of an IR excess. In this range our measure of the IR excess
is also consistently small, SN $\less$ 0.35.

Most of the stars that appear in Table~\ref{Tab:IR} are also marked in Fig.~\ref{Fig:IR}
(V Cen, AY Cen, S TrA, V1334 Cyg, $\zeta$ Gem, SU Cyg, $\eta$ Aql).
Polaris is not analysed in \citet{Trahin19}, but \cite{Merand2006} quote $\alpha \sim 0.016$~mag also based on the SPIPS method.
The value for SN is also small, and therefore the IR excess in $\alpha$ UMi is probably not significant.
The second star is $\delta$ Cep with $\alpha = 0.06$~mag. In this case the IR excess might be real.
The value for SN is likely to be underestimated in this work because of the relatively large error bar on the effective temperature in this particular case.

The three stars in Table~\ref{Tab:IR} for which no significant IR excess is found have been analysed by \citet{Trahin19}: 
$\beta$ Dor ($\alpha = 0.08$), RS Pup ($\alpha = 0.04$), and $l$ Car ($\alpha = 0.04$~mag).
The case of $\beta$ Dor is the most puzzling as the value of $\alpha$ appears significant.
On the other hand only a relatively poor LRS spectrum is available which is less constraining than the IRS spectra.
For the other two stars, the value of $\alpha$ is small and overall consistent with my finding of no excess.
In earlier works on $l$ Car by the same group \citet{Kervella2006} reported an IR excess similar to \citet{Trahin19},
but \citet{Breitfelder16} found no excess. Again, this points to the difficulty of establishing small levels of IR excess with confidence.

The analysis also revealed possible IR excess in stars that do not have an MIR spectrum available, namely
LS Pup (confirmed by \citealt{Trahin19} as well), and the stars with
SN $>$ 0.4, AD Cru, EX Cyg, XX Vel (the three most likely cases),
and the more uncertain cases of V5567 Sgr, CR Cep, FN Vel, DF Lac, and HW Car.
The cepheid ID 2 is also marked in the figure. The analysis of the IR excess was done using the standard value of the interstellar reddening.
If the analysis were repeated fixing the effective temperature to the spectroscopic one, or one consistent with
the location of the IS (see Sect.~\ref{S-IND}), the excess would disappear.
It indicates that a very wrong choice of the reddening could lead one to believe that there were an IR excess.

\subsection{Period-luminosity and period-radius relations}
\label{S-PL}

The final topic to be discussed are the relations between period and bolometric luminosity and radius.
In a first selection outliers in the HRD are excluded. As discussed, this could be due to a misclassification of the object,
an incorrect distance (that is, luminosity), an incorrect reddening, or a combination of these effects.
Improved distances ({\it Gaia} DR3), improved 3D reddening models, and improved light curve classification
(possibly from end-of-mission {\it Gaia} light curves or light curves from other time-domain surveys) may in the future
shed light on why exactly some stars are outliers in the present analysis.

\begin{figure}
\centering

\begin{minipage}{0.49\textwidth}
\resizebox{\hsize}{!}{\includegraphics{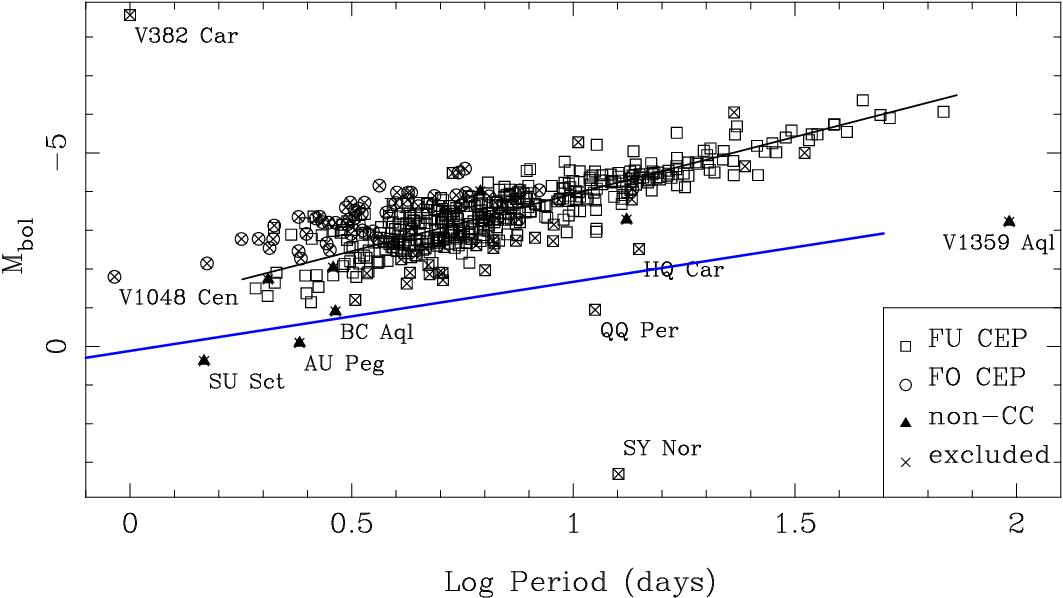}}
\end{minipage}
\begin{minipage}{0.49\textwidth}
\resizebox{\hsize}{!}{\includegraphics{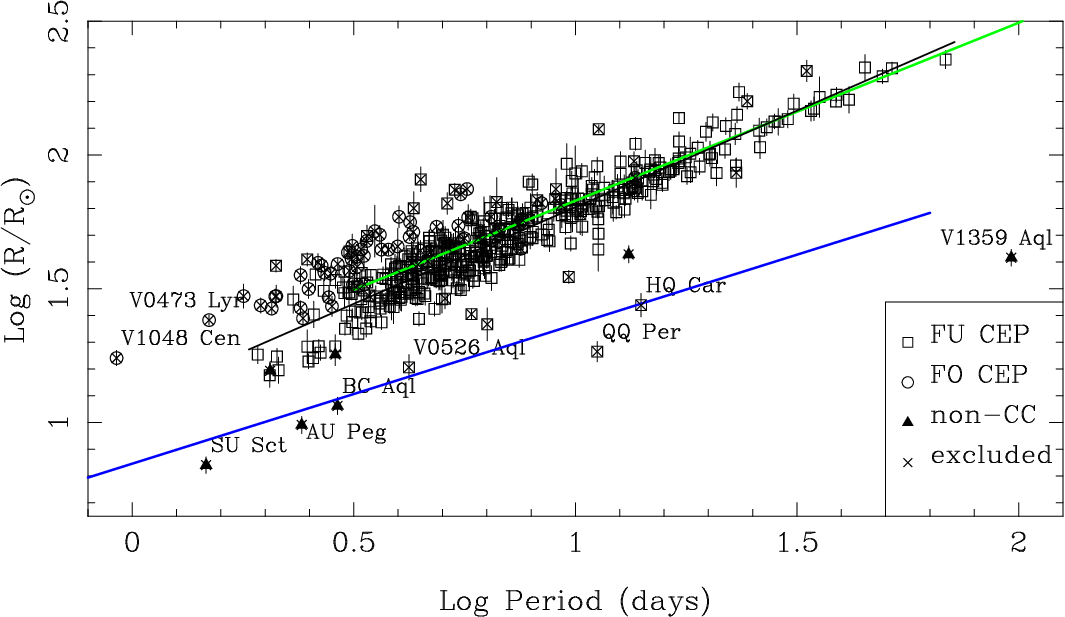}}
\end{minipage}

\caption{%
Top panel shows the period-$M_{\rm bol}$ relation.
Error bars in $M_{\rm bol}$ are not show as they are smaller than the symbol size.
Some outlying stars are identified.
The black line indicates the best fit to the fundamental mode cepheids (excluding outliers).
The blue line gives the $PL$ relation for T2Cs in the MCs from \citet{GrJu17b}.
The bottom panel shows the period-radius relation.
Some outlying stars are identified.
The black line indicates the best fit to the fundamental mode cepheids (excluding outliers).
The blue line gives the $PR$ relation for T2Cs in the MCs from \citet{GrJu17b}.
The green line is the theoretical $PR$ relation from \citet{Anderson16}.
}
\label{Fig:PLPR}
\end{figure}

Figure~\ref{Fig:PLPR} shows the results.
To be included in the fitting, objects are selected to have
$L < 50000$~\lsol, $T_{\rm eff} < 7000$~K, $T_{\rm eff} > 4000$~K, and ($L < 350$~\lsol\ or $\log L > (-12.083 \cdot \log(T_{\rm eff}) +47.5)$).
The last relation is a line across the HRD that eliminates the stars that are much cooler than expected from the bulk of stars and the
red-edge of the IS.
The weighted linear least-squares fits are done using only FU mode pulsators
(i.e. the period of overtone pulsators are not `fundamentalised' and included in the fit), and
iterative 3$\sigma$ clipping.
The best fit relations are
\begin{equation}
M_{\rm bol} =  (-2.95 \pm 0.09) \log P  + (-0.98 \pm 0.07),
\end{equation}
using 380 stars and with an rms of 0.40~mag, and
\begin{equation}
\log R =  (0.721 \pm 0.013) \log P  + (1.083 \pm 0.012),
\end{equation}
using 372 stars and with an rms of 0.067~dex.

This empirical $PR$ relation is based on the largest sample of Galactic cepheids.
It agrees largely with previous estimates (see the compilation in Table~\ref{Tab:PR}), although many $PR$-relations
  are based on the Baade-Wesselink method that depends on the adopted projection ($p$) factor that converts radial
  velocity to pulsational velocity.
  Theoretical $PR$ relations tend to give slightly shallower slopes,
but the maximum difference  with the recent work of \citet{Anderson16} (average rotation value, $Z= 0.014$,
averaged over 2nd and 3rd crossing; the green line in Fig.~\ref{Fig:PLPR}) is only +13\% at $\log P= 0.5$, and $-5$\% at  $\log P= 1.8$.
  These relations have been derived from the luminosities based on the adopted distance.
  Relations that also take into account an estimated error on the distance in the luminosities are
  given in Appendix~\ref{AppDist}, but the effect is small.

What is noticeable is that a number of stars scatter around the $PL$ and $PR$ relation for T2Cs derived
for the MCs \citep{GrJu17b}. Some stars were already known to be T2C (SU Sct, AU Peg, BC Aql) but others
have on occasion also been classified as CCs but are clearly T2C (e.g. QQ Per, HQ Car).

\begin{table*}
\setlength{\tabcolsep}{1.6mm}
\caption{Period-Radius relations of the form $\log R =  a \log P  + b$}
\begin{tabular}{ccrrl} \hline \hline 
     $a$          &        $b$        &  N   & rms   & Reference and remarks \\
\hline 
$0.721 \pm 0.013$ & $1.083 \pm 0.012$ & 372 & 0.057 & this work \\
$0.665          $ & $1.164          $  &   - &  -    & \citet{Anderson16}, theory, $Z= 0.014$, $\omega_{\rm ini} = 0.5$, average 2nd and 3rd crossing \\
$0.676 \pm 0.006$ & $1.173 \pm 0.008$ &   - &  -    & \cite {Petroni03}, theory, solar metallicity \\ 
$0.684 \pm 0.007$ & $1.135 \pm 0.002$ &   - & 0.020 & \cite {Gallenne17}, $p= 1.33 -0.08 \log P$ \\ %
$0.665 \pm 0.012$ & $1.136 \pm 0.014$ & 162 & 0.055 & \cite {Gr13}, $p= 1.50 -0.24 \log P$ \\ 
$0.737          $ & $1.074          $ & 162 &  -    & \cite {Gr13}, $p= 1.33$ \\ 
$0.75  \pm 0.03 $ & $1.10  \pm 0.03 $ &  26 & 0.036 & \cite {Molinaro11}, $p= 1.27$ \\ 
$0.767 \pm 0.009$ & $1.091 \pm 0.011$ &   8 &  -    & \citet{Kervella04PR} \\ %
$0.747 \pm 0.028$ & $1.071 \pm 0.025$ &  13 & 0.009 & \citet{Turner02}, $p= 1.31$ \\ 
$0.680 \pm 0.017$ & $1.146 \pm 0.007$ &  44 & 0.045 & \citet{Gieren99}, $p= 1.39 -0.03 \log P$ \\
$0.750 \pm 0.024$ & $1.075 \pm 0.007$ &  28 & 0.036 & \citet{Gieren98}, $p= 1.39 -0.03 \log P$ \\
$0.751 \pm 0.026$ & $1.070 \pm 0.008$ &  40 & 0.051 & \citet{LS95}, $p= 1.36$ \\ 
\hline
\end{tabular} 
\label{Tab:PR}
\end{table*}

\section{Discussion and summary}
\label{S-Dis}

The spectral energy distributions representative of mean light of 477 cepheids were constructed and modelled with stellar photospheres
(and a dust component in some cases).
Using distances and reddenings from the literature, this resulted in estimates of the bolometric luminosity and effective temperature at mean light,
which ultimately allow for the derivation of the period-luminosity and period-radius relations based on a sample of more than 370 fundamental-mode classical cepheids.

The average positions of the stars in the HRD are largely consistent with the theoretical ISs for CCs or T2Cs.
About 5\% of the stars in the sample are outliers in the sense that they are significantly cooler or hotter than expected.
The likely cause in at least a fraction of the stars is the degeneracy between the fitting of the effective temperature and the adopted reddening.
In cases when multiple effective temperature determinations from spectroscopy exist so that an accurate mean effective temperature can be determined
this mean temperature is in better agreement with that expected from the IS. This suggests that the photometric $T_{\rm eff}$ is biased by an incorrect reddening.

Two 3D reddening models (STILISM, \citealt{Lallement18}, and from \citealt{Green2019}) have been used to compare
the values to the adopted reddening values (see the Appendix).
There are systematic differences between these two models of order 15\%.
Compared to the adopted reddenings from the literature the \citet{Green2019} model shows a lower dispersion than the STILISM model,
but it is limited to stars north of declination $-30\degr$.
This uncertainty introduces an additional uncertainty in the derived parameters, in particular, luminosity and effective temperature.
  Some test calculations show that the uncertainty in the adoped $E(B-V)$ in Table~\ref{Tab-Targets} could lead to changes
  in luminosity, but that these are smaller than the quoted error or smaller than 0.15$L$ in 80\% of the sample.

This paper, like G18, is written with the tremendous potential offered by {\it Gaia} in mind.
Future data releases will provide information that will impact and improve on the results obtained here.
Primarily improved parallaxes, taking into account binarity in the astrometrical solution.
This impacts the cepheids, but also the 3D reddening models that use  {\it Gaia} parallaxes as input.
Secondly, improved lightcurves that will allow an homogeneous an improved classification of variable stars and of their subtypes.
Thirdly, astrophysical parameters derived from the Bp, Rp, and RVS spectra, in particular effective temperatures (but also metallicity or reddening).

One of the main topics addressed in this paper is the IR excess around CCs which is self-consistently modelled assuming
a circumstellar dust shell.
First the stars with MIR spectra are analysed (see Table~\ref{Tab:IR}) and then the entire sample.
The results for the stars with a (likely) IR excess (SN $>$ 0.4) are shown in Table~\ref{Tab:IRex}.
The first six stars ($\zeta$ Gem to V1334 Cyg also appear Table~\ref{Tab:IR}). The only difference in the fitting is
that the slope of the density law is fixed $p = 2$ while previously is was also a free parameter (and the dust mixture is fixed to 80\% iron).
The other parameters do change noticeably, although formally mostly within the error bars. The results of having $p$ fixed
is that the dust optical depth and dust temperature at the inner radius are better constrained.
A comparison of the results for the stars with spectra and those where the IR excess is only determined from photometry (LS Pup to HW Car)
shows how important the spectra are in constraining the fit. The results on the dust parameters are much less constrained.
Obtaining flux-calibrated MIR spectra for these ten stars would be valuable in confirming and better constraining the nature of the IR
excess.

Although statistically significantly better fits can be obtained by including a circumstellar dust shell the question remains if this
is a physically correct interpretation. The stars where $T_{\rm inn}$ is determined with a signal-to-noise better than three have
values around 800~K (3 stars), and then ranging from 1000 to 2000~K (five stars). Condensation temperatures of corundum, forsterite and metallic iron
are around 1670, 1350, and 1360~K, respectively, at a pressure typical of the Solar nebula \citep{Lodders03}.
At lower pressures these temperatures are lower. The 50\% condensation temperature of metallic iron condensing on pre-existing
corundum grains is about 1250~K at $10^{-5}$ bar and about 970~K at $10^{-10}$ bar \citep{Tachibana11}. 
Extrapolating the data in Fig.~5 in \citet{Grossman72} suggests condensation temperatures well below 1000~K for metallic iron as well as enstatite and forsterite at low pressures.
So at least for some of the stars the high values of the temperature at the inner radius appear to be in conflict with the expected condensation temperature.
  In addition, interferometric observations have resolved the CSE around a few CCs ($l$ Car, \citealt{Kervella2006}; Polaris, \citealt{Merand2006})
  that show the emission originates from 2-3\rstar, where temperatures are larger than the condensation temperatature (see the link between
  radius and dust temperature at the inner radius in Tables~\ref{Tab:IR} and \ref{Tab:IRex}).

When this paper was ready for submission, a work by \citet{Hocde19} appeared also discussing the IR excess around CCs.
They claim that the excess can not be explained by a hot or cold dust shell, and show that a thin shell of ionised gas
is able to explain the observations. They investigated RS Pup, $\zeta$ Gem, $\eta$ Aql, V Cen, SU Cyg and use the SPIPS method, paying
special attention to the analysis of the {\it Spitzer} IRS spectra. They modelled the IR excess using DUSTY \citep{Ivezic_D}
and used two extreme dust models; pure silicates and pure iron dust.
The silicate model is ruled out immediately because of the absence of the 9.7~\mum\ feature, while the iron dust model is ruled out
because the $T_{\rm inn}$ they find in their calculation (2238~K) to achieve a good fit is much larger than the condensation temperature.
They then proceed to show that a thin shell of ionised gas can explain the IR excess around these five stars.

Unfortunately, \citet{Hocde19} did not try mixtures of dust species which might be key.
Interestingly, although pure iron dust was also considered in the fitting of the twelve stars with MIR spectra, it never
turned out to be the best fit, but rather the 80, 90, or 94\% mixtures that were tried (Table~\ref{Tab:IR}).
Nevertheless the condensation temperatures, condensation sequence and nucleation and dust growth under the low-density conditions
expected in such a hypothetical CSE are a serious concern
and the proposed thin shell of ionised gas by \citet{Hocde19} is an interesting and viable alternative to explain the IR excess.

\begin{table*}
\caption{Stars with IR excess from figure~\ref{Fig:IR}.}
\begin{tabular}{rrrrrrrrrrrrrrrrrl} \hline \hline 
  Name    &    $T_{\rm eff}$ & $L$      & $\sigma$ &     $\tau$  & $\sigma$ & $T_{\rm inn}$  &  $\sigma$ &  $R_{\rm inn}$ & MLR      \\
          &    (K)         & (\lsol)  &   (\lsol) &            &          &     (K)        &    (K)    & ($R_{\star}$)   & (\msolyr)      \\
\hline 
$\zeta$ Gem   &  5375*  &   3399.2  &    3.36  &  2.64e-04  &  2.72e-05  &  1023.6  &    50.5  &    47.3  &  7.1e-10  \\ 
$\eta$ Aql    &  5500  &   2977.1  &    3.35  &  8.66e-04  &  5.83e-05  &  1673.6  &    56.5  &    11.2  &  4.9e-10  \\ 
AY Cen     &  5625  &   1822.3  &    4.40  &  3.54e-03  &  7.30e-04  &  1984.8  &   164.0  &     7.2  &  9.7e-10  \\ 
SU Cyg     &  6250*  &   1077.9  &    1.38  &  4.12e-04  &  5.91e-05  &  1285.2  &    66.4  &    33.5  &  3.3e-10  \\ 
V Cen      &  5625*  &   1720.9  &    2.05  &  3.46e-04  &  3.95e-05  &   772.6  &    38.9  &   121.9  &  1.6e-09  \\ 
V1334 Cyg  &  6000*  &   1985.7  &    1.88  &  1.48e-04  &  1.20e-05  &   750.4  &    28.6  &   154.1  &  8.0e-10  \\ 
LS Pup     &  5625*  &   4341.7  &  150.35  &  1.52e-01  &  9.35e-02  &  2972.8  &  1009.6  &     2.4  &  2.1e-08  \\ 
AD Cru     &  5625*  &   1822.7  &   61.64  &  4.14e-02  &  4.20e-02  &  1738.2  &   695.5  &    10.6  &  1.7e-08  \\ 
EX Cyg     &  5625  &   1034.1  &   27.47  &  6.64e-03  &  4.25e-02  &  1181.9  &  2738.2  &    34.0  &  6.5e-09  \\ 
S TrA      &  5750*  &    504.1  &    0.91  &  6.06e-04  &  8.19e-05  &   893.3  &    47.9  &    83.6  &  9.7e-10  \\ 
XX Vel     &  5625*  &   2710.5  &   61.39  &  4.86e-02  &  3.70e-02  &  2268.5  &   700.3  &     4.9  &  1.1e-08  \\ 
V5567 Sgr  &  5625*  &   1784.5  &   39.24  &  6.30e-03  &  1.30e-02  &  1609.4  &  1231.2  &    13.3  &  3.2e-09  \\ 
CR Cep     &  5375*  &   1666.7  &   49.38  &  1.45e-03  &  2.08e-03  &   486.3  &   509.4  &   376.4  &  2.2e-08  \\ 
FN Vel     &  5625  &   1183.1  &   20.38  &  5.89e-03  &  5.02e-03  &  1629.5  &   822.4  &    12.8  &  2.3e-09  \\ 
DF Lac     &  5750  &   1460.9  &   28.98  &  1.28e-02  &  2.03e-02  &  1853.8  &  1163.3  &     9.2  &  3.9e-09  \\ 
HW Car     &  5125  &   2478.4  &   61.13  &  1.88e-02  &  2.76e-02  &  2609.6  &  2231.1  &     2.7  &  2.7e-09  \\ 
\hline
\end{tabular} 
\label{Tab:IRex}
\tablefoot{
Column~1 gives the name of the cepheid.
Column~2 gives the effective temperature. A `*' after the temperature indicates that it changed w.r.t. the standard model.
Columns~3-4 give the luminosity and error bar.
Columns~5-6 give the dust optical depth at 0.55~$\mu$m and error bar.
Columns~7-8 give the temperature at the inner radius and error bar.
All error bars have been scaled to a reduced $\chi^2$ of unity.
Column~9 gives the mass-loss rate, assuming a dust-to-gas ratio of 1/200, and an expansion velocity of 200~\ks.
The values should be used with extreme caution as they are uncertain by a factor of ten due to the uncertainties in the
adopted expansion velocity, dust-to-gas ratio, dust opacity and the modelling itself.
}  
\end{table*}

\begin{acknowledgements}
I would like to thank Dr. Jan Lub for stressing the importance of including Walraven photometry,
Dr. Alexandre Gallenne for providing the mid-IR spectra of T Mon and X Sgr in electronic format, 
and Dr. Monika Jurkovic for interesting discussions during her visit to the ROB.
This work has made use of data from the European Space Agency (ESA) mission {\it Gaia} 
(\url{http://www.cosmos.esa.int/gaia}), processed by the {\it Gaia} Data Processing and Analysis Consortium 
(DPAC, \url{http://www.cosmos.esa.int/web/gaia/dpac/consortium}). 
Funding for the DPAC has been provided by national institutions, in particular
the institutions participating in the {\it Gaia} Multilateral Agreement.
This research is based on observations with AKARI, a JAXA project with the participation of ESA.
This research has made use of the SIMBAD database and the VizieR catalogue access tool 
operated at CDS, Strasbourg, France.
\end{acknowledgements}

\bibliographystyle{aa.bst}
\bibliography{references.bib}

\begin{appendix}

\section{Different reddening values}
\label{AppRedd}

\begin{figure}
\centering

\begin{minipage}{0.38\textwidth}
\resizebox{\hsize}{!}{\includegraphics{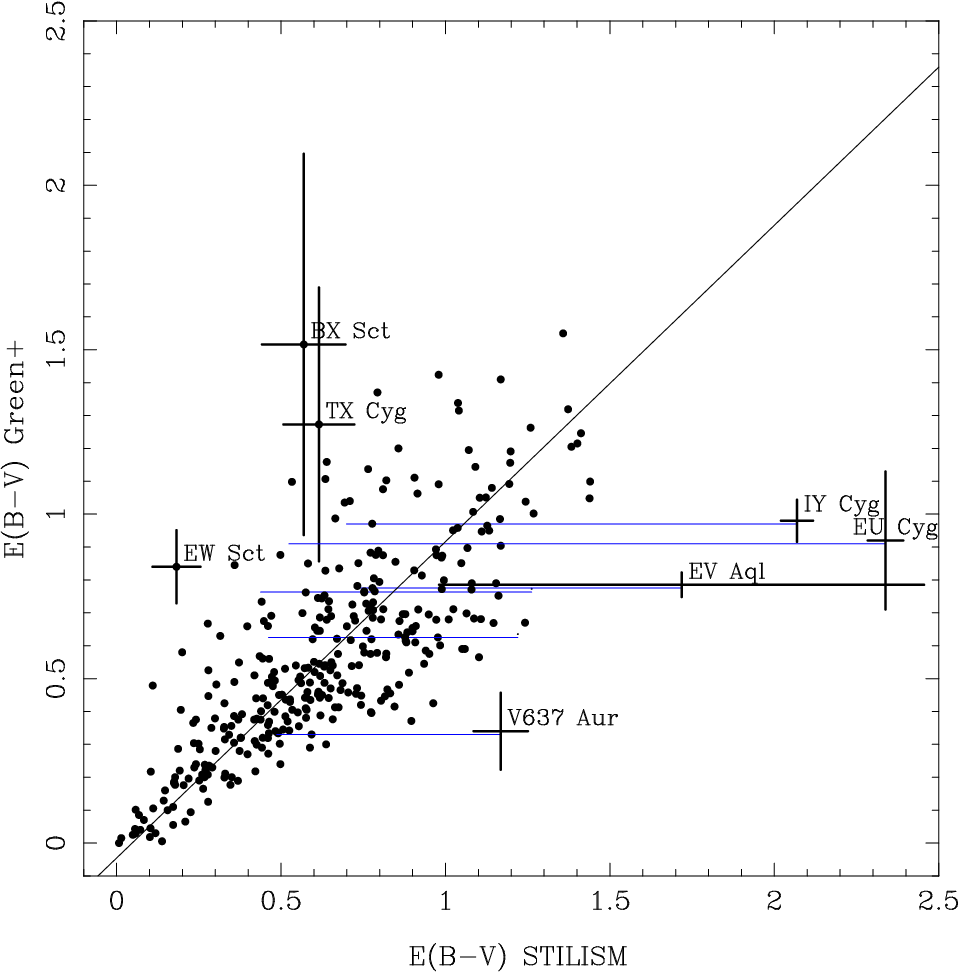}}
\end{minipage}

\begin{minipage}{0.38\textwidth}
\resizebox{\hsize}{!}{\includegraphics{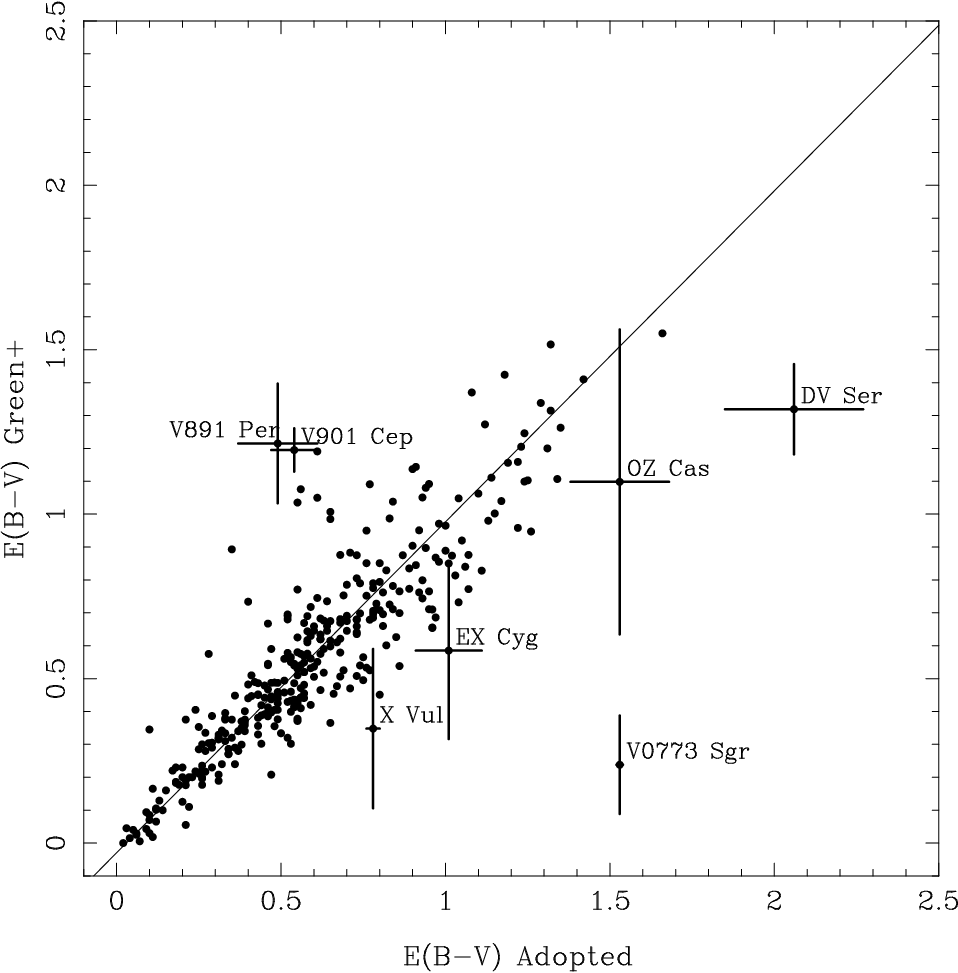}}
\end{minipage}

\begin{minipage}{0.38\textwidth}
\resizebox{\hsize}{!}{\includegraphics{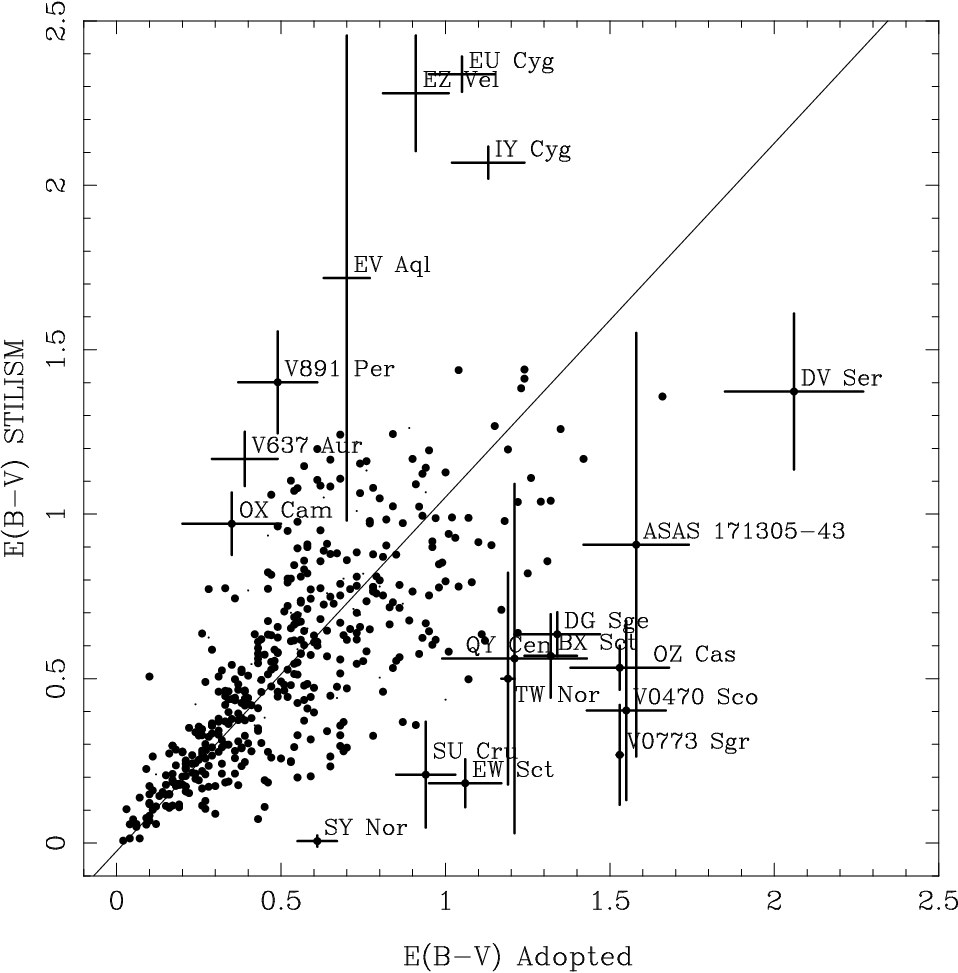}}
\end{minipage}

\caption{
Different $E(B-V)$ values plotted against each other. Outliers are marked and plotted with error bars.
In the top panel some of the stars are connected by a blue line. The left point indicates the reddening
at the maximum distance available in the STLISM grid in that particular direction. The right point indicates the reddening
estimated at the distance of the cepheid by a linear extrapolation (see Sect.~\ref{S-EBV} for details).
The solid lines indicate the least-squares fir from Table~\ref{App:Fit}.
}
\label{Fig:App}
\end{figure}

As outlined in the main text, the choice of reddening in the fitting procedure can have an important impact on the results.
This was shown for the CCs in the inner disk which have values of $A_{\rm V}$  in the range 7-16.
Other indications were stars where the spectroscopic effective temperatures differed significantly from the ones determined from the SED fitting and outliers in the HRD.
Table~\ref{Tab:App} collects the different reddening values, and
Fig.~\ref{Fig:App} shows the different values plotted against each other.
Outliers are marked and plotted with error bars.
In the top panel outliers witch large reddening from STILISM are likely due to an inappropriate extrapolation
from the the reddening at the maximum distance available in the STLISM grid in that particular direction to the
reddening at the distance of the cepheid (see Sect.~\ref{S-EBV} for details).

Linear bi-sector fits were made in all cases (excluding the marked outliers), as well as determining the median and median-absolute-deviation (MAD)
of the ratio of the quantities (in the sense y-axis value/x-axis value). The fit results are collected in Table~\ref{App:Fit}. 
There are systematic differences between the three sets of reddenings (in the sense STILISM $>$ adopted values $>$ Green et al.)
of order 5-15\%.

\longtab[1]{
\setlength{\tabcolsep}{1.4mm}
\begin{longtable}{lllllllllll}
\caption{Different reddening values} \\
\hline  \hline
   Name   &  $E(B-V)$ & $\sigma$ & $E(B-V)$ & $\sigma$ & $E(B-V)$ & $E(B-V)$ & $\sigma$ & $E(B-V)$ & $\sigma$ & Remarks \\  
\hline
\endfirsthead
\caption{continued.}\\
\hline\hline
   Name   &  $E(B-V)$ & $\sigma$ & $E(B-V)$ & $\sigma$ & $E(B-V)$ & $E(B-V)$ & $\sigma$ & $E(B-V)$ & $\sigma$ & Remarks \\  
\hline
\endhead
\hline
\endfoot
   AA Gem          & 0.340 & 0.040 & -     & -     & 0.327 & 0.461 & 0.054 & 0.272 & 0.057 &  \\ 
   AA Mon          & 0.770 & 0.020 & -     & -     & 0.628 & 0.972 & 0.093 & 0.679 & 0.087 &  \\ 
   AB Cam          & 0.620 & 0.040 & -     & -     & 0.800 & 0.951 & 0.105 & 0.575 & 0.204 &  \\ 
   AC Mon          & 0.510 & 0.030 & -     & -     & 0.464 & 0.464 & 0.144 & 0.560 & 0.314 &  \\ 
   AD Cam          & 0.870 & 0.010 & -     & -     & 0.830 & 0.973 & 0.058 & 0.875 & 0.110 &  \\ 
   AD Cru          & 0.640 & 0.010 & -     & -     & 0.426 & 0.568 & 0.081 & -     & -     &  \\ 
   AD Gem          & 0.210 & 0.050 & -     & -     & 0.141 & 0.172 & 0.028 & 0.055 & 0.044 &  \\ 
   AD Pup          & 0.360 & 0.020 & -     & -     & 0.257 & 0.498 & 0.084 & 0.240 & 0.078 &  \\
\end{longtable}
\tablefoot{
Cols.~2-3 are the adopted reddening and error, repeated from Table~\ref{Tab-Targets},
Cols.~4-5 are the reddening and error from \citet{Trahin19},
Cols.~6-8 are the results from the 3D model by \cite{Lallement18}.
  The $E(B-V)$ value in Col.~6 is the one in the available grid at the distance closest to distance of the object.
  The value in Col.~7 is the $E(B-V)$ value at the distance of the object, based on the extrapolation explained in the main text.
  Column~8 lists the error.
Cols.~9-10 are the results from the 3D model by \citet{Green2019},
The last column indicates if a star is an outlier in the HRD (Fig.~\ref{Fig:HRD}, labelled as HRD)
or in the comparison with the spectroscopic temperature determinations (Fig.~\ref{Fig:Teff}, labelled TEF).
}
}

\begin{table*} 
\caption{Relations between different reddening values.}

\begin{tabular}{cccccc} \hline \hline 
 Comparison &  \multicolumn{3}{c}{$y = a \cdot x + b$}          &   \multicolumn{2}{c}{$y/x$}  \\
            &     $a$      & $b$                       & rms    & median & MAD       \\
 \\
\hline 

Green et al. versus STILISM & 0.962  & $-0.046$  & 0.18  & 0.855  & 0.23  \\ 
Green et al. versus adopted & 1.007  & $-0.030$  & 0.13  & 0.930  & 0.14 \\ 
STILISM      versus adopted & 1.077  & $-0.025$  & 0.20  & 1.038  & 0.22  \\ 
\hline

\end{tabular} 
\tablefoot{
Columns~2-4 give the coefficients of a linear bi-sector fit, and the rms.
Columns~5-6 give the median and the median-absolute-deviation of the ratio of the two quantities.
}
\label{App:Fit}
\end{table*}

\section{Additional fits and plots}
\label{AppDist}

\begin{figure}
\centering

\begin{minipage}{0.49\textwidth}
\resizebox{\hsize}{!}{\includegraphics{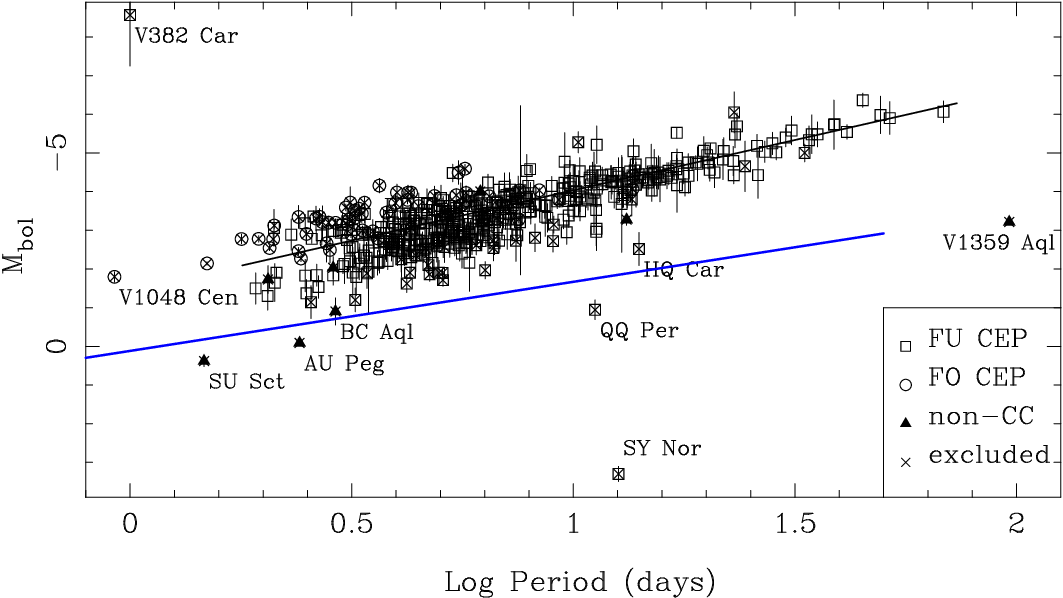}}
\end{minipage}
\begin{minipage}{0.49\textwidth}
\resizebox{\hsize}{!}{\includegraphics{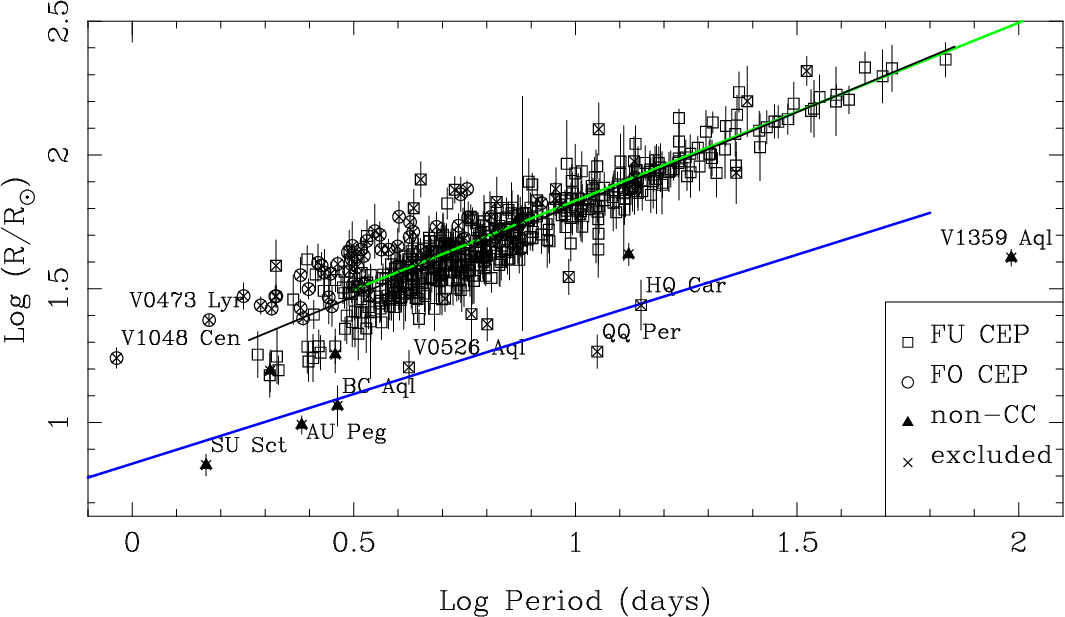}}
\end{minipage}

\caption{%
  As Fig.~\ref{Fig:PLPR}, but error bars in $M_{\rm bol}$ are now plotted.
}
\label{Fig:PLPRdist}
\end{figure}

Figure~\ref{Fig:PLPRdist} shows the period-luminosity and period-radius relations when the error in the distance
is taken into account in the error estimate of the luminosity.
The adopted error in the distance is listed in Col.~8 of Table~\ref{Tab-Targets}.
For the CCs in the inner disk they are based on I19, otherwise on the error estimates given in \cite{BJ18}.
If upper and lower error bars were given the geometric mean was taken.
For stars were the GDR2 parallax clearly is in error (and thus also the value in \cite{BJ18}) the value
from G18 is taken ($\alpha$ UMi, $l$ Car, $\delta$ Cep, S Mus). For stars not in GDR2 (and therefore not considered in G18)
an error in distance was estimated from stars at very similar distances (RY Vel, V340 Nor, IY Cep).

The period-luminosity relation becomes
\begin{equation}
M_{\rm bol} =  (-2.60 \pm 0.07) \log P  + (-1.43 \pm 0.06), 
\end{equation}
using 380 stars and with an rms of 0.42~mag.
The difference with Eq.~1 at $P = 10$~d is 0.1~mag.

The period-radius relation becomes
\begin{equation}
\log R =  (0.689 \pm 0.014) \log P  + (1.126 \pm 0.012), 
\end{equation}
using 375 stars and with an rms of 0.072~dex.
The difference with Eq.~2 at $P = 10$~d is 0.011~dex (2.6\%).

\end{appendix}

\end{document}